\documentclass[aps,pra,reprint,amsmath,amssymb,superscriptaddress,longbibliography]{revtex4-1}

\usepackage{graphicx}
\usepackage[colorlinks=true,urlcolor=blue,citecolor=blue,linkcolor=blue,bookmarks=false,pdfstartview={FitH}]{hyperref}
\usepackage{soul}

\begin{document}

\title{Raman spectroscopy of $f$-electron metals: An example of CeB$_{6}$}

\author{Mai Ye}
\email{mye@physics.rutgers.edu}
\affiliation{Department of Physics and Astronomy, Rutgers University, Piscataway, NJ 08854, USA}
\author{H.-H. Kung}
\affiliation{Department of Physics and Astronomy, Rutgers University, Piscataway, NJ 08854, USA}
\author{Priscila F. S. Rosa}
\affiliation{Los Alamos National Laboratory, Los Alamos, NM 87545, USA}
\author{Eric D. Bauer}
\affiliation{Los Alamos National Laboratory, Los Alamos, NM 87545, USA}
\author{Zachary Fisk}
\affiliation{Department of Physics and Astronomy, University of California, Irvine, CA 92697, USA}
\author{Girsh Blumberg}
\email{girsh@physics.rutgers.edu}
\affiliation{Department of Physics and Astronomy, Rutgers University, Piscataway, NJ 08854, USA}
\affiliation{National Institute of Chemical Physics and Biophysics, 12618 Tallinn, Estonia}

\date{June 14, 2019}

\begin{abstract}
We performed an optical spectroscopy study of electronic and magnetic 
excitations for a rare-earth system with a single electron 
quasi-localized in the $f$-shell on an ion at high-symmetry 
crystallographic site in application to CeB$_{6}$ heavy-fermion 
metal. We carried out group-theoretical classification of the electronic crystal field (CF) transitions and assessed their coupling to light cross-sections for polarization resolved Raman scattering processes. We discuss applicability of symmetrized Raman susceptibility to studies of exotic charge and spin high multiplet ordering phases in $f$-electron systems. We study temperature effects on intra- and inter-multiplet CF transitions and also on the coupling between the CF excitations with the lattice vibrations. We acquired temperature dependence of the low-frequency polarization resolved Raman response for all Raman-allowed symmetry channels: A$_{1g}$, E$_{g}$, T$_{1g}$, and T$_{2g}$ of the cubic O$_{h}$ point group. We demonstrate that T$_{1g}$-symmetry static Raman susceptibility shows a temperature dependence which is consistent with the previously-reported magnetic susceptibility data. Such behavior in the T$_{1g}$ channel signifies the presence of long wavelength magnetic fluctuations, which is interpreted as a manifestation of ferromagnetic correlations induced by tendency towards quadrupolar ordering.
\end{abstract}

\doi{10.1103/PhysRevMaterials.3.065003}
\maketitle

\section{Introduction\label{sec:Intro}}
Strongly correlated $d$- and $f$-electron systems support a rich variety of low-temperature phases, including magnetism and superconductivity~\cite{Tokura1998,Dagotto2005,Moore2009,Pfleiderer2009}. Among these phases, long-range order of multipoles, namely high-rank electric or magnetic moments, has great interest~\cite{Kuramoto2009,Santini2009,Cameron2016,Suzuki2018}. For example, second-rank quadrupolar moments could lead to novel phenomena including the quadrupolar Kondo effect~\cite{Cox1987} and quadrupole-fluctuation-mediated superconductivity~\cite{Kotegawa2003}. In $d$-electron systems, the orbital angular momentum is usually quenched by large crystal-field (CF) splitting, hindering multipolar moments. $f$-electron systems, on the other hand, are suitable choices to study multipolar interactions and ordering phenomena by virtue of the interplay of the spin and orbital degrees of freedom. Indeed, the actinide dioxides, in which $5f$-electrons play an important role, serve as a paradigm for understanding the physics of multipolar interactions~\cite{Santini2009}. Quadrupolar orderings have also been discovered in a number of $4f$-electron compounds~\cite{Tm1993,Tm1998,Dy2000,Np2002,Pr2014,Cameron2016}.

\begin{figure}
\includegraphics[width=0.48\textwidth]{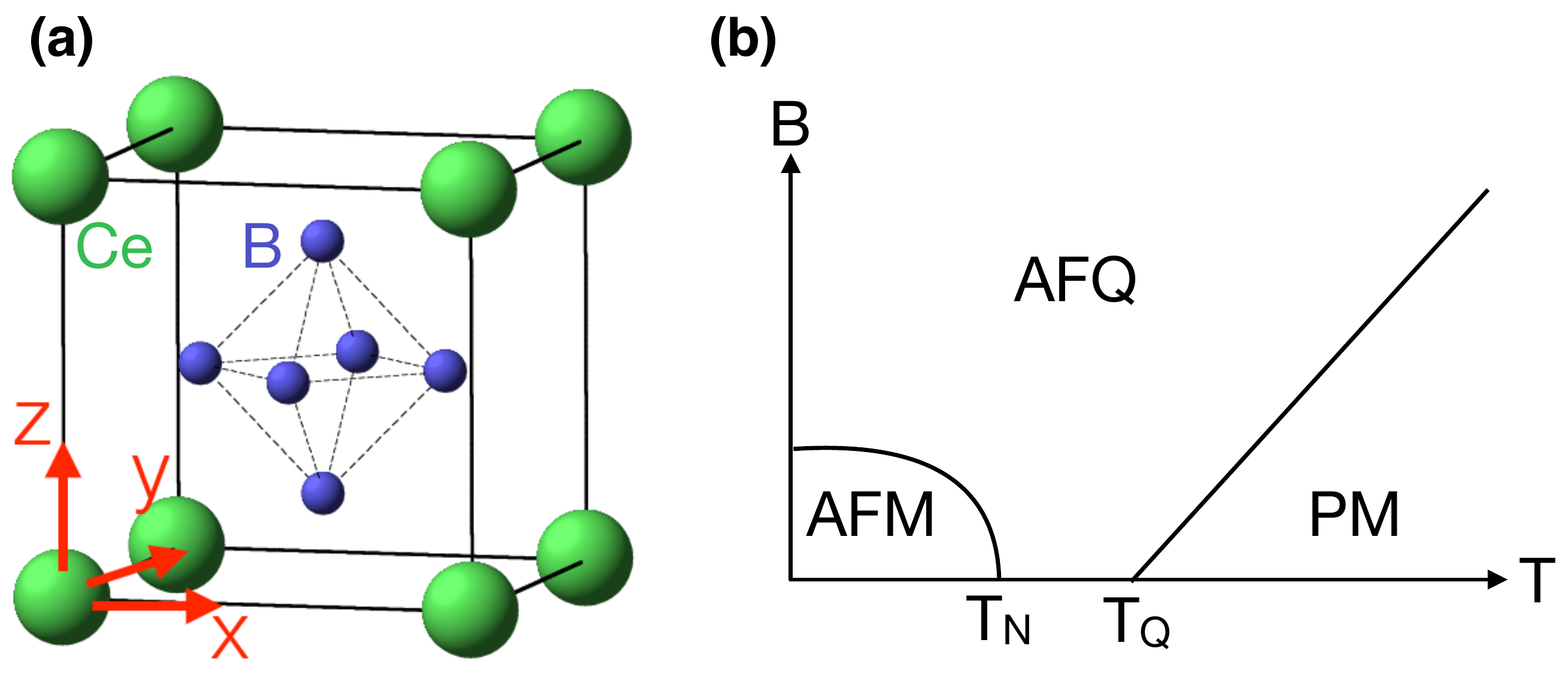}
\caption{\label{fig:Intro1}(a) Crystal structure of CeB$_{6}$. (b) A sketch of field-temperature phase diagram for CeB$_{6}$.}
\end{figure}

CeB$_{6}$, with its simple chemical composition, lattice structure, and electronic configuration, is considered a prototypical example of heavy-fermion metal with quadrupolar ordering. This material has a cubic structure (space group Pm$\overline{3}$m, No. 221; point group O$_{h}$) composed of cerium ions and boron octahedrons [Fig.~\ref{fig:Intro1}(a)]. Every Ce$^{3+}$ ion has only one electron in its $4f$ orbital and O$_{h}$ site symmetry. CeB$_{6}$ undergoes a second-order phase transition into a non-magnetic phase at T$_{Q}$\,=\,3.2\,K, before developing an antiferromagnetic (AFM) order below T$_{N}$\,=\,2.3\,K~\cite{Takase1980,Fujita1980}. The AFM phase has a double-Q commensurate magnetic structure with Q$_1$=(0.25, 0.25, 0) and Q$_2$=(0.25, 0.25, 0.5)~\cite{Burlet1982,Zaharko2003}. As for the non-magnetic phase, neutron scattering shows no structural transition at T$_{Q}$~\cite{Zaharko2003}. Resonant X-ray diffraction determines that this non-magnetic phase involves an orbital ordering with wavevector (0.5, 0.5, 0.5)~\cite{Nakao2001}, and the C$_{44}$ elastic constant, related to $\epsilon_{xy}$-type strains, shows an anomaly at T$_{Q}$~\cite{Nakamura1994}. Based on these results, it is generally believed that the non-magnetic phase is a two-sublattice arrangement of Ce$^{3+}$ $O_{xy}$-type electric quadrupole moments, with a wavevector (0.5, 0.5, 0.5)~\cite{Cameron2016}. This proposed antiferroquadrupolar (AFQ) model is consistent with experimental data in the presence of magnetic field~\cite{Hanzawa2000,Kunimori2012,Matsumura2009,Matsumura2012,Schlottmann2012}, but to our knowledge, up to now there is no direct evidence demonstrating the $O_{xy}$-type AFQ order in zero field. A sketch of field-temperature phase diagram for CeB$_{6}$ is shown in Fig.~\ref{fig:Intro1}(b). 

All experimental results reported in this study correspond to the zero-field paramagnetic (PM) phase, namely, the data is acquired at T\,$>$\,T$_{Q}$.

In the recent years multiple experimental studies have revealed the importance of unexpected ferromagnetic (FM) correlations in the low-temperature ordering phenomena of CeB$_{6}$. In the AFQ phase with finite magnetic field, electron spin resonance (ESR) with narrow linewidth was uncovered, pointing to existence of FM correlations~\cite{Demishev2009}. Theoretical study suggested that such FM correlations result from AFQ ordering~\cite{Schlottmann2012}. A zone-center excitation at the (110) point, following the energy of ESR, was found by inelastic neutron scattering (INS)~\cite{Portnichenko2016}. In the AFQ phase at zero magnetic field, this finite-energy mode collapses into a quasi-elastic peak~\cite{Jang2014}. Moreover, intense FM fluctuations were uncovered in the AFM phase, suggesting propensity to FM instability~\cite{Jang2014}.

\begin{figure}[t]
\includegraphics[width=0.48\textwidth]{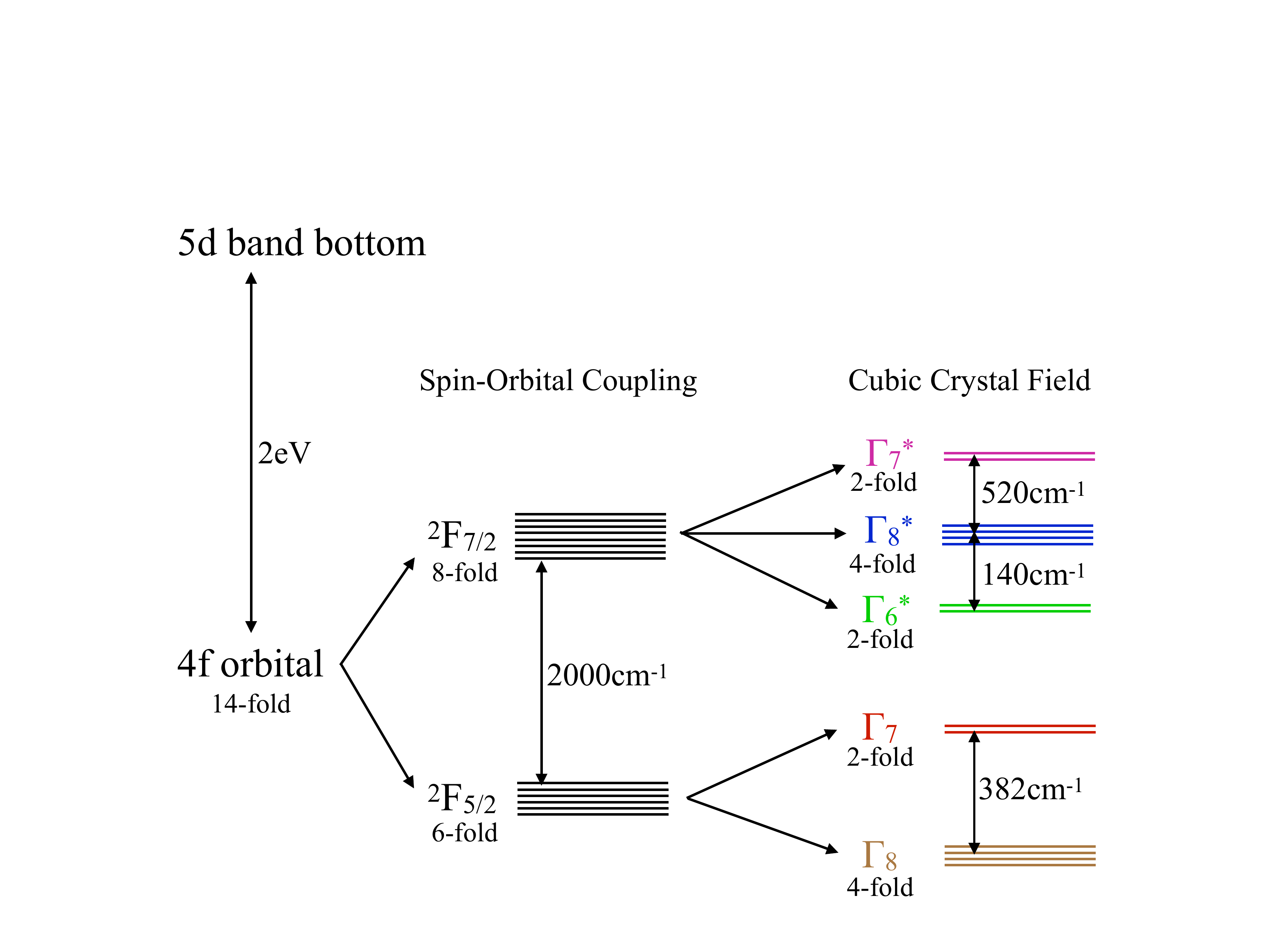}
\caption{\label{fig:Intro2}
Schematic energy diagram illustrating the splitting of 4f orbital by spin-orbital coupling and cubic crystal field. The same color scheme is used in Figs.~\ref{fig:Intro2},~\ref{fig:CF1},~\ref{fig:CF2}, and~\ref{fig:CF4} to identify the four crystal-field transitions.}     
\end{figure}

Both the AFQ and AFM phases are closely related to the CF ground state~\cite{Cameron2016}. In CeB$_{6}$, 6-fold degenerate $^2F_{5/2}$ is the ground multiplet, and 8-fold $^2F_{7/2}$ is the lowest-energy excited multiplet [Fig.~\ref{fig:Intro2}]. These two multiplets were identified in photoemission spectroscopy studies~\cite{Takahashi1995,Koitzsch2016} by the self-energy effects~\footnote{Photoemission spectroscopy probes energy states below the Fermi level, thus, ARPES cannot directly access the $^2F_{7/2}$ multiplet. However, virtual excitations to the narrow $^2F_{7/2}$ multiplet contribute to the self energy of the spectral function, making the $^2F_{7/2}$ multiplet identifiable in the photoemission spectra. Details of this mechanism can be found in Refs.~\cite{Gunnarsson1983,Coleman1984,Patthey1987}}. From group theory analysis~\cite{Koster1963}, the cubic CF potential splits the $^2F_{5/2}$ multiplet into quartet $\Gamma_8$ and doublet $\Gamma_7$ states, and the $^2F_{7/2}$ multiplet into doublet $\Gamma_6^*$, doublet $\Gamma_7^*$, and quartet $\Gamma_8^*$ 
states~\footnote{Asterisks are used to distinguish the CF states of the $^2F_{7/2}$ multiplet ($\Gamma_6^*$, $\Gamma_7^*$ and $\Gamma_8^*$) from those of the $^2F_{5/2}$ multiplet ($\Gamma_7$ and $\Gamma_8$).}. For the $^2F_{5/2}$ multiplet, the $\Gamma_8$ state is the ground state~\cite{Sato1984,Zirngiebl1984,Sakai1997,Sundermann2017} and the $\Gamma_7$ state has an energy of 372\,cm$^{-1}$ at room temperature~\cite{Zirngiebl1984,Loewenhaupt1985ccf}. For the $^2F_{7/2}$ multiplet, the energy of the CF levels has not been determined experimentally.

In order to better understand the low-temperature ordering phenomena 
in CeB$_{6}$, a more detailed study of the interplay of CF 
excitations, lattice dynamics and the FM correlations is required. 
Raman spectroscopy is a suitable technique providing 
symmetry-resolved excitation spectra of electronic, magnetic, and 
phononic degrees of freedom. As a photon-in-photon-out inelastic 
scattering process, polarization-resolved Raman scattering has the 
unique advantage of high energy-resolution and the ability to 
disentangle the excitation spectra into individual symmetry channels. 
The symmetry of a particular excitation can be identified by 
controlling the polarization of the incident and scattered 
light~\cite{Hayes2004}. 
This experimental method has been 
successfully used to study CF excitations 
\cite{Cardona2000,Guntherodt1987}; 
it is a well-fitted choice of 
investigating the intra- and inter-multiplet CF excitations of 
CeB$_{6}$. 
Moreover, Raman scattering makes it possible to study the 
excitations in the magnetic dipolar (T$_{1g}$ of O$_{h}$ group) and 
electric quadrupolar (E$_{g}$ and T$_{2g}$ of O$_{h}$ group) channels 
separately. Thus, the relationship between the quadrupolar 
correlations and FM correlations can be clarified. Notice that 
quadrupolar excitations involve a change of the component of angular 
momentum along the quantization axis by two quantum units. Among 
conventional experimental probes, only photons can induce quadrupolar 
excitations.                       

In this paper, we present a comprehensive study of CeB$_{6}$ using 
optical secondary-emission spectroscopy. We identify an intense 
photo-luminescence feature corresponding to $5d-4f$ recombination 
process. We analyze the temperature-dependence of both intra- and 
inter-multiplet CF excitations, and illustrate the interaction 
between light and CF states by a model Hamiltonian calculation. We 
draw information about the electron-phonon interaction by studying 
lattice dynamics. We observe dynamical magnetic fluctuations related 
to the ordered broken-symmetry states. Especially, we demonstrate two 
virtues of Raman scattering which have not been generally 
appreciated: first, the temperature dependence of the parameters of 
CF excitations reveals the interaction between $f$-electrons and 
itinerant electrons; and second, the low-energy Raman response probes 
dynamical fluctuations related to exotic multipolar 
ordering.                

The rest of this paper is organized as follows. In Sec.~\ref{sec:Exp} we describe the sample preparation and experimental setup. In Sec.~\ref{sec:Res} we present and discuss the experimental results; in this section, we first show an overview of the main spectral features in SubSec.~\ref{subsec:OV} and then discuss them separately in the following subsections. In SubSec.~\ref{subsec:PL} we show the high-energy photo-luminescence (CF) feature. In SubSec.~\ref{subsec:CF} we discuss the CF excitations. Specifically, in \ref{subsubsec:Iden} we present the four lowest-energy CF excitations of Ce$^{3+}$ ions, and identify the symmetry of the CF states; in \ref{subsubsec:Tem}, we analyze the temperature dependence of the CF parameters, and explain the observed anomaly on the basis of Kondo effect; in \ref{subsubsec:Model}, we build a single-ion Hamiltonian, and fit the measured CF energies with this Hamiltonian to evaluate the SOC and CF strength, and to obtain the wavefunctions of eigenstates. In SubSec.~\ref{subsec:P} we discuss lattice dynamics. The asymmetric lineshape, and relatively large full-width-at-half-maximum (FWHM) of the optical phonon modes point to electron-phonon interaction. In SubSec.~\ref{subsec:QE} we discuss quasi-elastic excitations. We find that quasi-elastic fluctuations in the symmetry channel containing magnetic excitations develops below 20\,K, and that the temperature dependence of the corresponding Raman susceptibility is consistent with the previously-reported static magnetic susceptibility data. Finally, in Sec.~\ref{sec:Con} we provide a summary of our 
observations and their implications. 

\section{Experimental\label{sec:Exp}}
Single crystals of CeB$_{6}$ were grown in Al flux by slow cooling from 1450\,$^\circ$C. The crystals were removed from the Al flux by leaching in NaOH solution~\cite{Foroozani2015,Canfield1992}. The sample measured in this study was cleaved in ambient condition to expose its (001) crystallographic plane; the cleaved surface was then examined under a Nomarski microscope to find a strain-free area.

\begin{table}[b]
\caption{\label{tab:Exp1}The relationship between the scattering geometries and the symmetry channels. Every scattering geometry is represented by E$_{i}$E$_{s}$, where E$_{i}$ and E$_{s}$ are the polarizations of incident and scattered light; X, Y, X' and Y' are the [100], [010], [110] and [1$\overline{1}$0] crystallographic directions; R and L are right and left circular polarizations. A$_{1g}$, E$_{g}$, T$_{1g}$ and T$_{2g}$ are the irreducible representations of the O$_{h}$ group.}
\begin{ruledtabular}
\begin{tabular}{ll}
Scattering Geometry&Symmetry Channel\\
\hline
XX&A$_{1g}$+4E$_g$\\
XY&T$_{1g}$+T$_{2g}$\\
X'X'&A$_{1g}$+E$_g$+T$_{2g}$\\
X'Y'&3E$_g$+T$_{1g}$\\
RR&A$_{1g}$+E$_g$+T$_{1g}$\\
RL&3E$_g$+T$_{2g}$\\
\end{tabular}
\end{ruledtabular}
\end{table}

Raman-scattering measurements were performed in a quasi-back scattering geometry from sample placed in a continuous helium-gas-flow cryostat. A set of lines from a Kr$^+$ ion laser, 476, 531, 647, 676 and 752\,nm, were used for excitation. Incident light with less than 10\,mW power was focused into a 50$\times$100\,$\mu$m$^{2}$ spot. The temperature points reported in this paper were corrected for laser heating, which was estimated to be 
0.5\,K$\slash$\,mW~\footnote{We mainly followed the procedure discussed in Ref.~\cite{Maksimov1992} to estimate the laser heating. The optical absorption coefficient data were extracted from the optical data~\cite{Kimura1990,Kimura1992,Kimura1994}, while the thermal conductivity data were taken from Ref.~\cite{Sera1996}.}.

Six polarization configurations were employed to probe excitations in different symmetry channels. The relationship between the scattering geometries and the symmetry channels~\cite{Hayes2004} is given in Table.~\ref{tab:Exp1}. The algebra used to decompose measured spectra into four symmetry channels is shown in Table.~\ref{tab:Exp2}. 

We used a custom triple-grating spectrometer with a liquid-nitrogen-cooled charge-coupled device (CCD) detector for analysis and collection of the scattered light. Low-resolution gratings with 150 lines per mm were used to measure the broad PL feature, while high-resolution gratings with 1800 lines per mm were used for measurements of the sharp Raman features. The data were corrected for the spectral response of the system.

For first-order scattering processes, the measured secondary-emission intensity $I(\omega,T)$ is related to the Raman response $\chi''(\omega,T)$ by $I(\omega,T)=[1+n(\omega,T)]\chi''(\omega,T)+L(\omega,T)$, where $n$ is the Bose factor, $\omega$ is excitation energy, $T$ with temperature, and $L(\omega,T)$ represents 
photo-luminescence~\cite{SM}. 
For the second-order acoustic-phonon scattering process to be 
discussed in SubSec.~\ref{subsec:P}, assuming the two constitute 
excitations have the same energy, $I(\omega,T)$ and 
$\chi''(\omega,T)$ are related by 
$I(\omega,T)=[1+n(\omega/2,T)]^2\chi''(\omega,T)+L(\omega,T)$~\cite{SM}. 

\begin{table}
\caption{\label{tab:Exp2}The algebra used in this study to decompose the data into four symmetry channels.}
\begin{ruledtabular}
\begin{tabular}{lc}
Symmetry Channel&Expression\\
\hline
A$_{1g}$&$(1/3)(XX+X'X'+RR-X'Y'-RL)$\\
E$_g$&$(1/6)(X'Y'+RL-XY)$\\
T$_{1g}$&$(1/2)(XY+RR-X'X')$\\
T$_{2g}$&$(1/2)(XY+RL-X'Y')$\\
\end{tabular}
\end{ruledtabular}
\end{table}

\section{Results and Discussion\label{sec:Res}}

\subsection{Overview\label{subsec:OV}}

\begin{figure}
\includegraphics[width=0.48\textwidth]{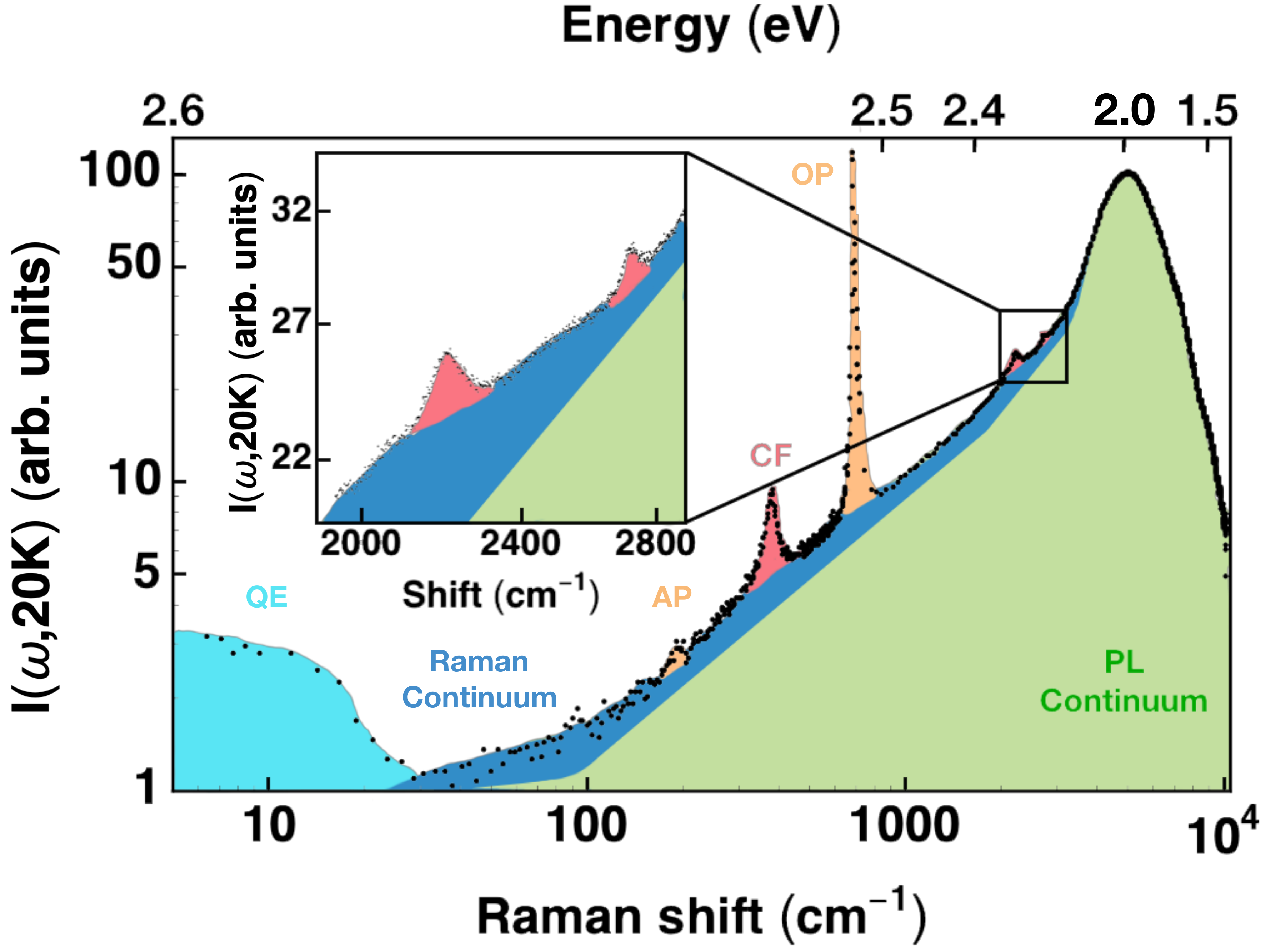}
\caption{\label{fig:OV} 
An overview of the low-temperature secondary-emission intensity measured in XY geometry at 20\,K with 476\,nm excitation in log-log scale. The top scale is the absolute energy of the secondary-emission photons in electron-Volts. The bottom scale is the energy loss, the laser-photon energy minus the scattered-photon energy, also called the Raman shift, in spectroscopic units cm$^{-1}$. The Raman features are superposed on a strong photo-luminescence continuum. Different Raman features are schematically represented by different colors: cyan, quasi-elastic (QE) Raman excitations; blue, the continuum of electronic Raman excitations; orange, second-order acoustic-phonon (AP) excitations and first-order optical-phonon (OP) excitations; red, crystal-field (CF) excitations; and green: the continuum of the photo-luminescence (PL).}      
\end{figure}

In Fig.~\ref{fig:OV} we present a typical secondary-emission spectrum over a large energy range, covering Raman features of distinct origins. Among the Raman features, quasi-elastic excitations have the lowest-energy. Second-order acoustic phonon excitations are at around 200\,cm$^{-1}$, while first-order optical phonon excitations are near 1000\,cm$^{-1}$. The energy of the intra-multiplet CF excitation is around 400\,cm$^{-1}$, while that of the inter-multiplet CF excitations is more than 2000\,cm$^{-1}$. The PL continuum arises from a broad PL peak at around 2.0\,eV. In the following subsections we will discuss every spectral feature separately in details.

\subsection{Photo-Luminescence\label{subsec:PL}}

\begin{figure}
\includegraphics[width=0.44\textwidth]{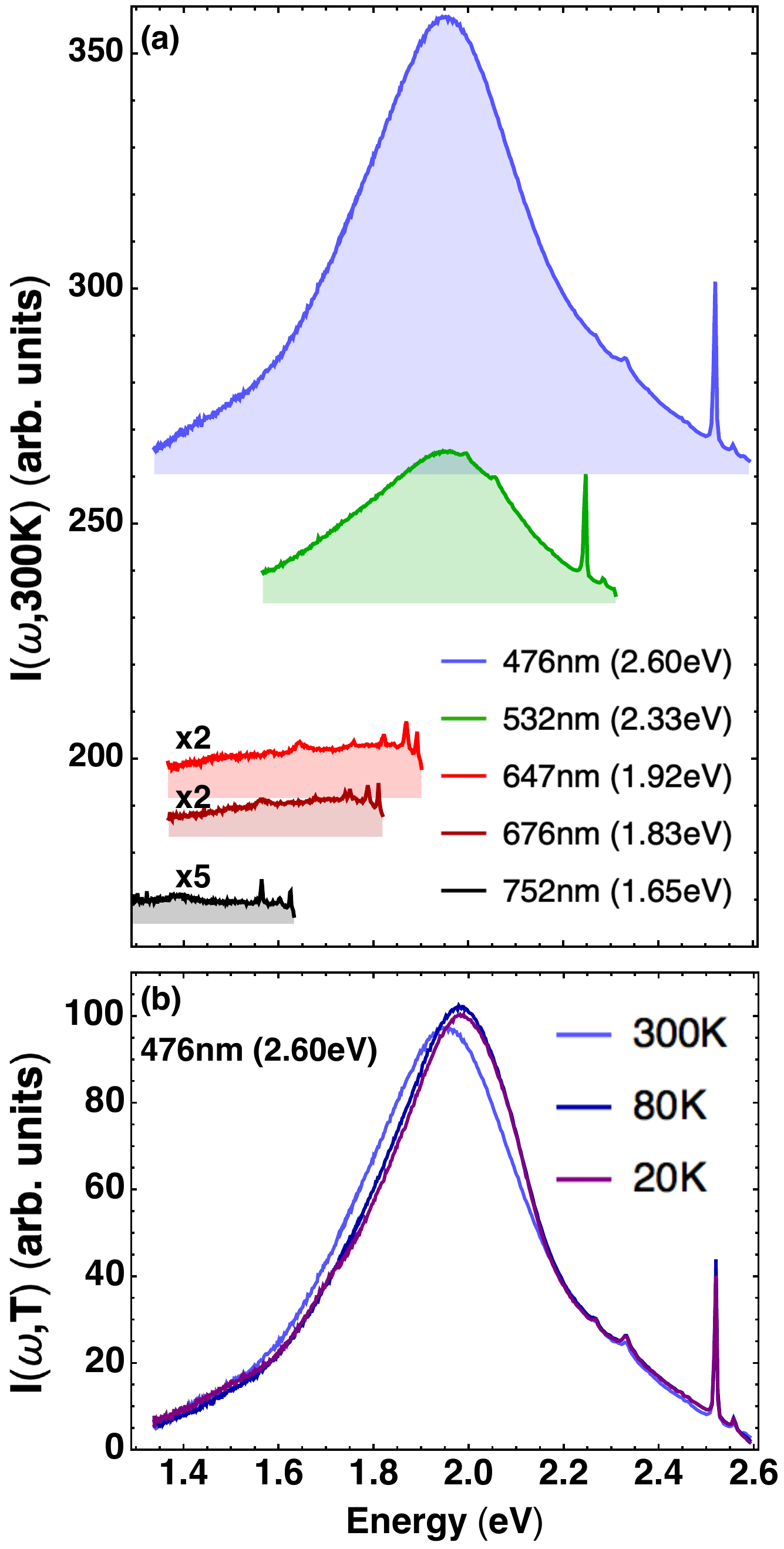}
\caption{\label{fig:PL1}
(a) Excitation dependence of the secondary-emission intensity I($\omega$,300\,K) measured in XY geometry at 300\,K. For clarity, each spectrum is vertically shifted by a factor proportional to the excitation energy. The broad peak which does not change in the absolute emission energy with excitation energy is a photo-luminescence feature, while the sharp modes which follow the excitation energy are the Raman features. (b) Temperature dependence of the photo-luminescence feature measured in XY geometry with 476\,nm excitation.}
\end{figure}

In Fig.~\ref{fig:PL1}(a) we show the excitation dependence of the PL feature at room temperature. The PL peak has 2.0\,eV excitation threshold, and excitations below 2.0\,eV threshold show predominantly Raman features. The PL feature is centered at 1.95\,eV, just below the threshold energy, and has about 0.4\,eV full width at half maximum (FWHM). Upon cooling the peak shifts slightly to higher energy [Fig.~\ref{fig:PL1}(b)].

The optical conductivity shows a shoulder at around 2.0\,eV~\cite{Kimura1990,Kimura1992,Kimura1994}, suggesting an optical gap. Band-structure calculations further indicate a 2.0\,eV gap between the Ce dispersive $5d$-band bottom and flat $4f$-band~\cite{Kitamura1994,Suvasini1996,Neupane2015}. We therefore attribute the PL peak to the recombination of the electron-hole excitations between the $5d$- and $4f$-bands. Transitions between $d$- and $f$-states are dipole allowed, and the energy separation of the $5d$-band bottom and the $4f$-band is consistent with the energy of this PL peak. The enhancement of PL intensity for excitations above the 2\,eV threshold results from the increase of the density of states (DOS) for the $4f$ to $5d$ interband transition.

\subsection{Crystal-Field Excitations\label{subsec:CF}}

\subsubsection{Identification\label{subsubsec:Iden}}

\begin{figure}
\includegraphics[width=0.48\textwidth]{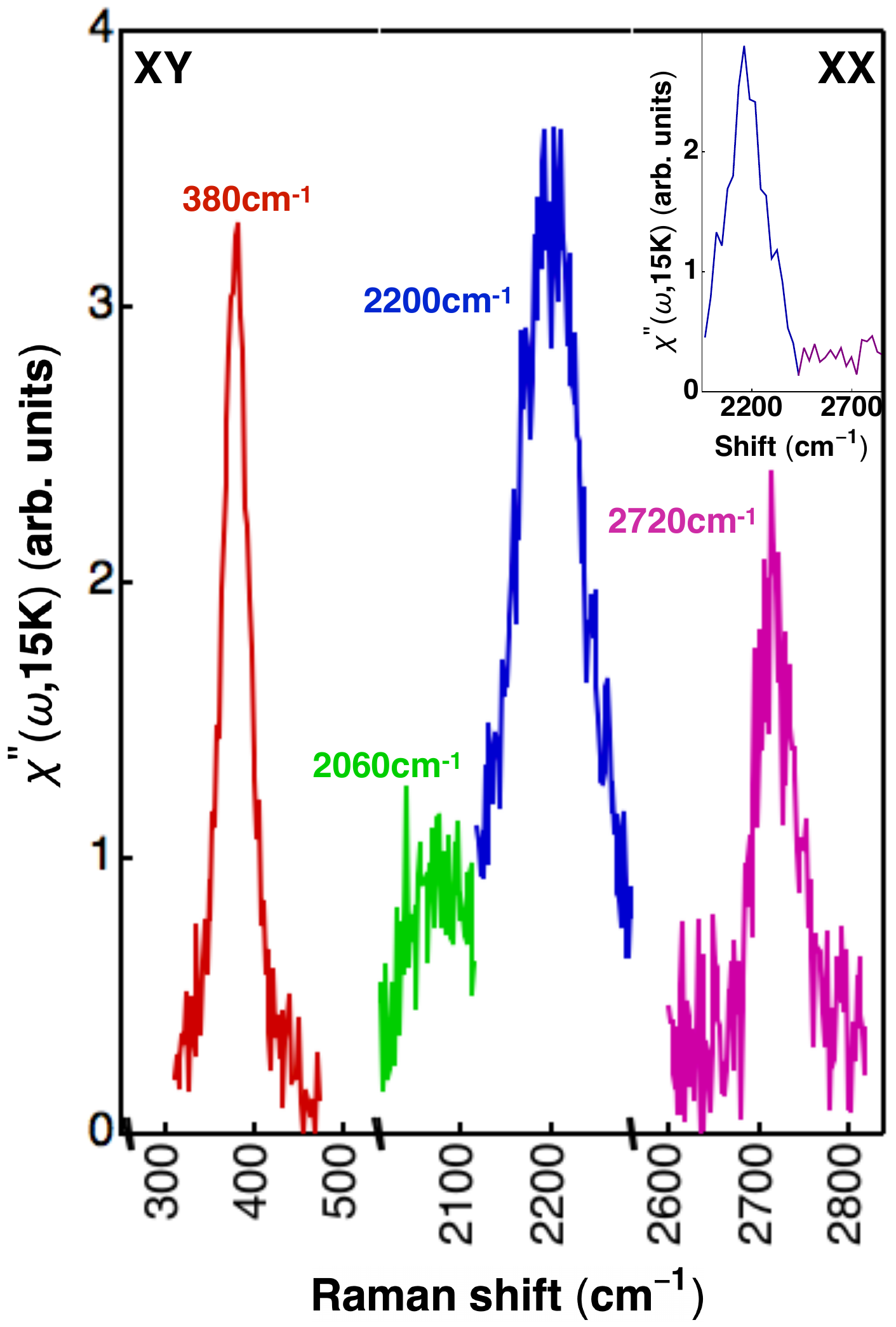}
\caption{\label{fig:CF1}Raman response $\chi^{\prime\prime}$($\omega$,15\,K) of the CF excitations measured in XY scattering geometry (T$_{1g}$+T$_{2g}$) with 476\,nm excitation at 15\,K. Three axis breakers are used on the horizontal axis in order to show the four excitations together. The spectral resolution is 3.5\,cm$^{-1}$. Inset: $\chi^{\prime\prime}$($\omega$,15\,K) measured in XX scattering geometry (A$_{1g}$+4E$_g$) at 15\,K. The spectral resolution of the inset is about 30\,cm$^{-1}$.}
\end{figure}

In total, there are four CF excitations from the $\Gamma_8$ ground state to the higher states within the $^2F_{5/2}$ and $^2F_{7/2}$ multiplets: one intra-multiplet excitation and three inter-multiplet excitations [Fig.~\ref{fig:Intro2}]. In Fig.~\ref{fig:CF1} we present the spectrum of the four CF excitations measured at 15\,K. Four peaks at 380\,cm$^{-1}$, 2060\,cm$^{-1}$, 2200\,cm$^{-1}$ and 2720\,cm$^{-1}$ are observed. The 380\,cm$^{-1}$ excitation is the intra-multiplet $\Gamma_8\rightarrow\Gamma_7$ transition. Among the three inter-multiplet excitations, only the $\Gamma_8\rightarrow\Gamma_8^*$ transition can have a finite $A_{1g}$ component~\cite{Koster1963}. In the inset of Fig.~\ref{fig:CF1} we show that among the inter-multiplet excitations only the one at 2200\,cm$^{-1}$ contains an $A_{1g}$ component. The 2200\,cm$^{-1}$ excitation is therefore assigned to the $\Gamma_8\rightarrow\Gamma_8^*$ transition. The CF excitation at 2720\,cm$^{-1}$, in turn, can only be a transition between the $\Gamma_8$ ground state and the $\Gamma_6^*$ or $\Gamma_7^*$ states. Raman scattering cannot distinguish between $\Gamma_8\rightarrow\Gamma_6^*$ and $\Gamma_8\rightarrow\Gamma_7^*$ transitions because they both contain the same irreducible representations~\cite{Koster1963}: $\Gamma_8\otimes\Gamma_6^*$\,=\,$\Gamma_8\otimes\Gamma_7^*$\,=\,$E_{g}\oplus T_{1g}\oplus T_{2g}$. However, we will show in \ref{subsubsec:Model} that the electron-cloud distribution of the $\Gamma_6^*$ state has the smallest overlap with the boron octahedrons, the $\Gamma_8^*$ state has intermediate overlap, and the $\Gamma_7^*$ state has the largest overlap. Because of the Coulomb repulsion between cerium and boron electrons, the $\Gamma_7^*$ state has the highest energy while the $\Gamma_6^*$ state has the lowest energy. Indeed, within the $^2F_{5/2}$ multiplet because the $\Gamma_7$ state has more overlap with the boron octahedrons it has a higher energy than the $\Gamma_8$ state. Therefore, the 2720\,cm$^{-1}$ excitation is assigned to the $\Gamma_8\rightarrow\Gamma_7^*$ transition, and the 2060\,cm$^{-1}$ excitation is assigned to the $\Gamma_8\rightarrow\Gamma_6^*$ transition.

\subsubsection{Temperature Dependence\label{subsubsec:Tem}}

\begin{figure}
\includegraphics[width=0.48\textwidth]{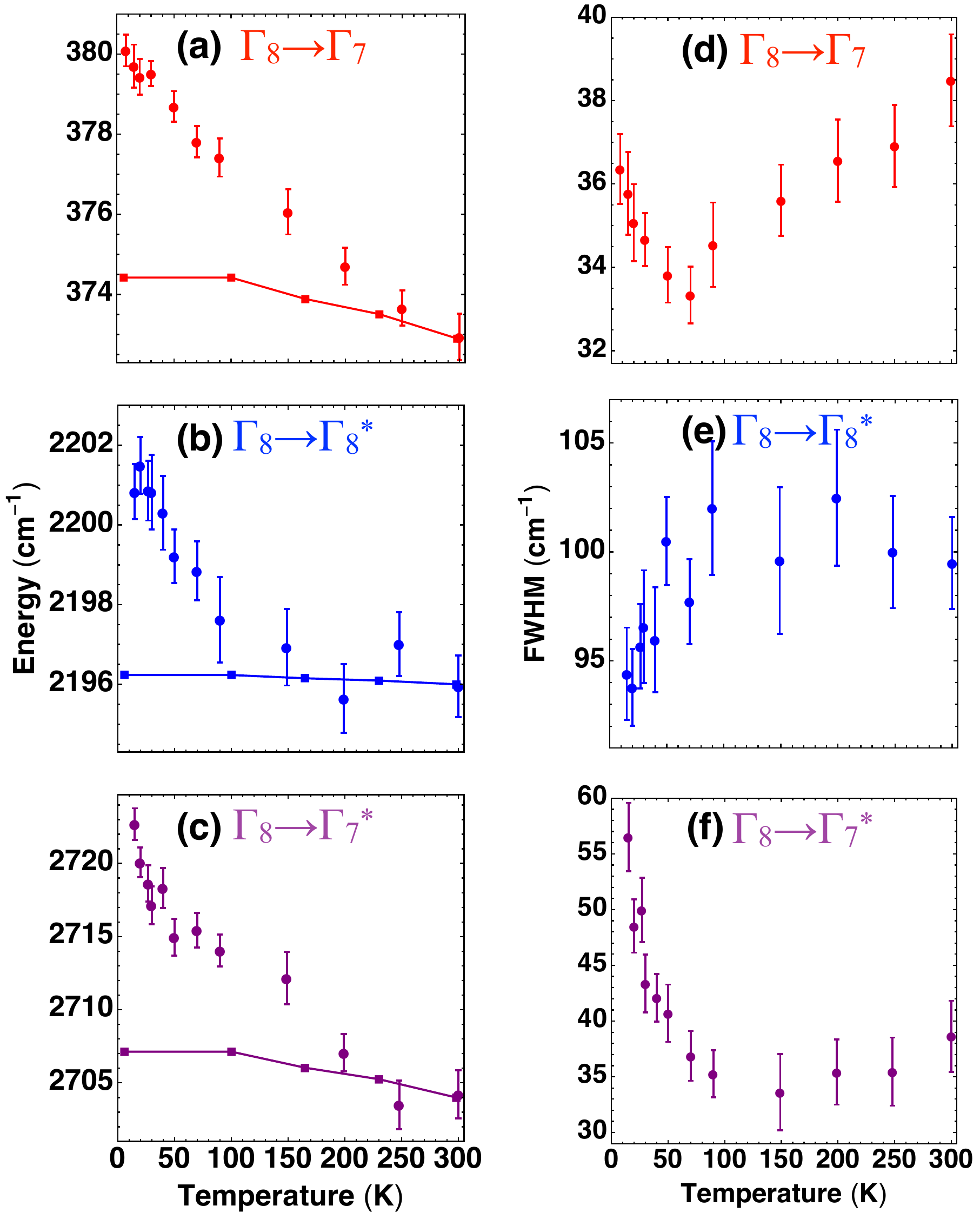}
\caption{\label{fig:CF2} Temperature dependence of the energy (a-c) and FWHM (d-f) of the $\Gamma_8\rightarrow\Gamma_7$, $\Gamma_8\rightarrow\Gamma_8^*$ and $\Gamma_8\rightarrow\Gamma_7^*$ CF excitations shown in Fig.~\ref{fig:CF1}. The line-joined square labels in (a-c) represent the excitation energies calculated by our model Hamiltonian calculation. The error bars represent one standard deviation of the Lorentzian fit.}
\end{figure}

In Fig.~\ref{fig:CF2} we present the temperature dependence of the energy and FWHM of three CF excitations. The spectral parameters of the CF excitations were obtained by fitting the measured spectral peaks with a Lorentzian lineshape.

On cooling from 304\,K to 15\,K the lattice contraction strengthens the electrostatic potential at the Ce sites resulting in increase of the $\Gamma_8\rightarrow\Gamma_7$, $\Gamma_8\rightarrow\Gamma_8^*$, and $\Gamma_8\rightarrow\Gamma_7^*$ transition energy by 7\,cm$^{-1}$, 5\,cm$^{-1}$, and 18\,cm$^{-1}$ respectively~\footnote{The energy of the $\Gamma_8\rightarrow\Gamma_7$ transition shows monotonic temperature dependence. The earlier reports, Ref.~\cite{Zirngiebl1984,Loewenhaupt1985ccf}, showed no temperature dependence of the transition energy on cooling from 300\,K to 20\,K, followed by a rapidly hardening on cooling below 20\,K.}. A discussion of the change of the energy of the CF states with increasing CF potential will be given in \ref{subsubsec:Model}.             

At room temperature, the CF spectral lines of CeB$_{6}$ are broader than those measured from Ce$^{3+}$ ions embedded in insulators, e.g. Ce-doped Y$_2$O$_3$~\cite{Nolas1994} or Ce-doped LuPO$_4$~\cite{Williams1989}. The broadening could be caused by two factors: first, lattice of Ce$^{3+}$ ions leads to small dispersion of the narrow $4f$-bands; and second, hopping of conduction electrons among the boron sites induces fluctuations of the electrostatic potential at the Ce sites, which broadens FWHM.

On cooling, the FWHM of the $\Gamma_8\rightarrow\Gamma_7$ and $\Gamma_8\rightarrow\Gamma_7^*$ CF transitions decrease from 300\,K to 80\,K, but anomalously increases below 80\,K [Fig.~\ref{fig:CF2}~(d) and (f)]. The decrease of FWHM is expected because lattice vibrations, causing fluctuations of the electrostatic potential at Ce sites, diminish with cooing. In order to understand the anomalous increase of FWHM below 80\,K, it is important to notice that the electrical resistivity of CeB$_{6}$ has its local minimum at 80\,K. The resistivity upturn below 80\,K results from the Kondo effect~\cite{Takase1980} due to increase in the rate of conduction electron scattering from the local moments at the Ce sites on cooling~\cite{Hewson1993,Amusia2015}. The Kondo effect shortens the lifetime of the $\Gamma_7$ and $\Gamma_7^*$ CF states, so the FWHM of the $\Gamma_8\rightarrow\Gamma_7$ and $\Gamma_8\rightarrow\Gamma_7^*$ CF transitions increases below 80\,K. Nevertheless, the FWHM of the $\Gamma_8\rightarrow\Gamma_8^*$ CF transition does not show an upturn below 80\,K [Fig.~\ref{fig:CF2}~(e)]. This is because the $\Gamma_8^*$ state has smaller overlap with the boron octahedrons than the $\Gamma_7$ and $\Gamma_7^*$ states, therefore, it is less influenced by the increased conduction electron scattering rate. 

Our data do not display directly the splitting of the $\Gamma_8$ CF ground state. However, the minimum FWHM of the $\Gamma_8\rightarrow\Gamma_7$ is around 33\,cm$^{-1}$ [Fig.~\ref{fig:CF2}~(d)]: if the splitting of the CF ground state is small, it would not be resolved. The previous studies suggested a splitting of 20\,cm$^{-1}$~\cite{Zirngiebl1984,Terzioglu2001}, which does not contradict our data.

\subsubsection{Model Hamiltonian Calculation\label{subsubsec:Model}}

To shed light on the nature of the CF transitions, we perform a model Hamiltonian calculation. We use the following single-ion Hamiltonian
\begin{equation}
H=E_0+H_{SOC}+H_{CF}~.
\label{eq:H}
\end{equation}
The first term $E_0$ represents the energy of unperturbed $4f$ shell. The value $E_0$ is chosen to put the $\Gamma_8$ ground state at zero energy.
The second term
\begin{equation}
H_{SOC}=\xi\hat{\mathbf{L}}\cdot\hat{\mathbf{\sigma}}
\label{eq:HSOC}
\end{equation}
describes the effect of SOC. Here $\xi$ is the SOC coefficient, $\hat{\mathbf{L}}$ is the orbital angular momentum operator and $\hat{\mathbf{\sigma}}$ are Pauli matrices.
The third term
\begin{equation}
H_{CF}=B_4(\hat{O}^0_4+5\hat{O}^4_4)+B_6(\hat{O}^0_6-21\hat{O}^4_6)
\label{eq:HCF}
\end{equation}
is the general expression for a CF potential of cubic site symmetry~\cite{Lea1962}, where $\hat{O}^0_4$, $\hat{O}^4_4$, $\hat{O}^0_6$ and $\hat{O}^4_6$ are Stevens operators~\cite{Stevens1952}, and $B_4$ and $B_6$ are the CF coefficients~\footnote{When comparing the CF coefficients across different literature, additional constants are needed~\cite{Hutchings1964,Kassman1970}.}:    
\begin{equation}
B_4=A_4<r^4>\beta~,
\label{eq:B4}
\end{equation}
\begin{equation}
B_6=A_6<r^6>\gamma~, 
\label{eq:B6}
\end{equation}
$A_4$ and $A_6$ are the geometrical coordination factors determined by the charge configuration around the Ce sites. Regardless of the specific configuration, $A_4\,\sim\,a^{-5}$ and $A_6\,\sim\,a^{-7}$, where $a$ is the lattice constant; $<r^4>$ and $<r^6>$ are the mean fourth and sixth powers of the radii of the Ce$^{3+}$ $4f$-orbital, and $\beta$ and $\gamma$ are the Stevens multiplicative factors~\cite{Stevens1952}.

\begin{figure*}
\includegraphics[width=0.98\textwidth]{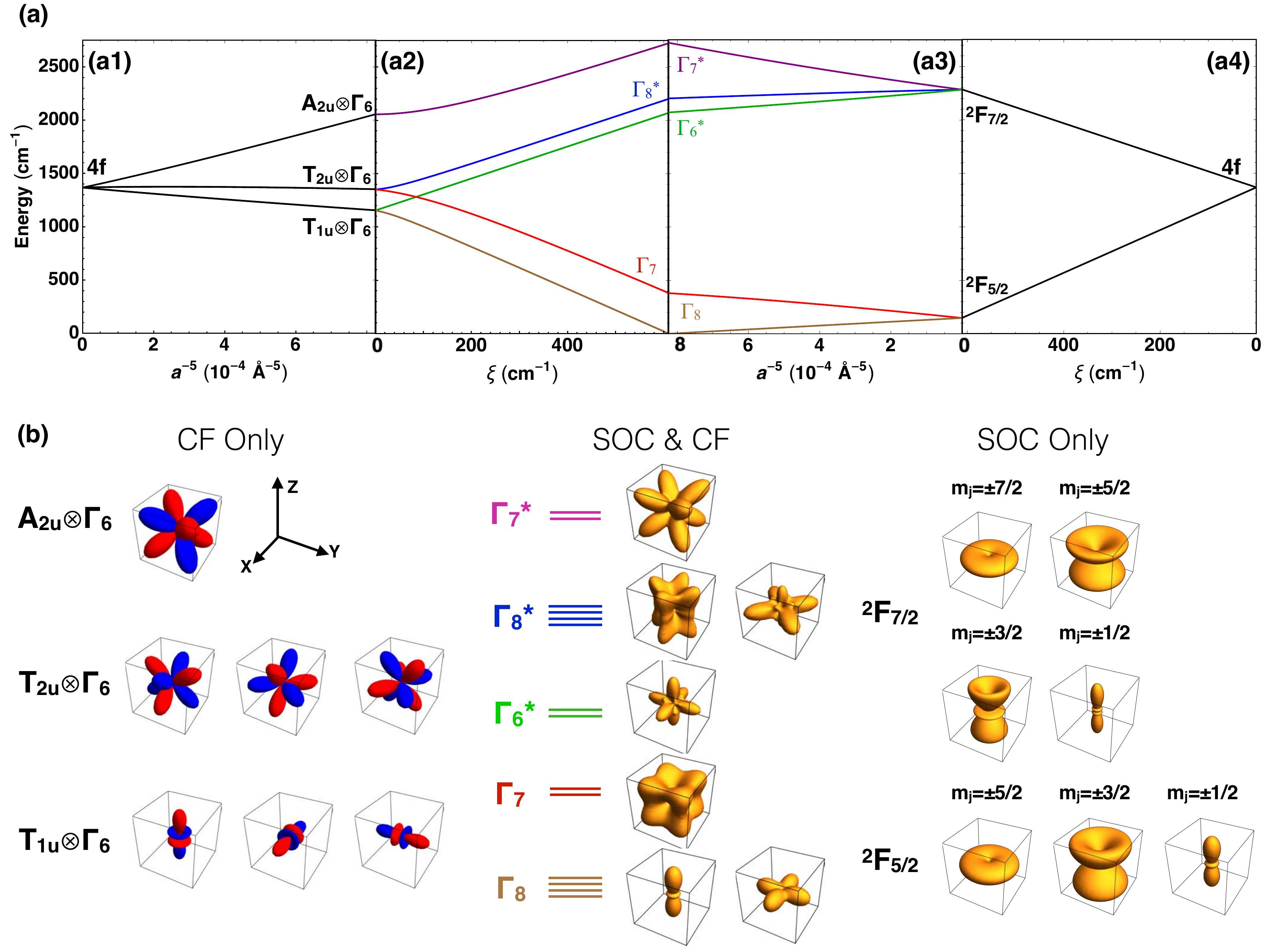}
\caption{\label{fig:CF4}Eigenenergies and eigenstates derived from the model Hamiltonian calculation. (a) Evolution of the $4f$-orbital energy with CF potential and SOC strength. [from left to center] Increasing CF potential in the absence of SOC (a1), and then increasing SOC strength in the existence of full CF potential (a2); [from left to center] increasing SOC in the absence of CF potential (a3), and then increasing CF potential in the existence of full SOC (a4). In this panel, the full SOC strength is $\xi$=610\,cm$^{-1}$, and the full CF potential strengths are $B_4$=-0.758\,cm$^{-1}$ and $B_6$=-0.0165\,cm$^{-1}$. (b) The wavefunctions and the angular electron-cloud distribution of the eigenstates. [left] The wavefunctions of the eigenstates when only CF potential is present. Red denotes positive value while blue denotes negative value; [middle] the angular electron-cloud distribution of the eigenstates when both SOC and CF potential are present; [right] the angular electron-cloud distribution of the eigenstates when only SOC is present.}
\end{figure*}

The effects of SOC and CF potential on the energy and angular electron-cloud distribution of the CF levels are illustrated in Fig.~\ref{fig:CF4}. In the absence of the SOC, the CF eigenfunctions could be classified by the irreducible representations (IRs) of O$_h$ double group. The relevant IRs are the one-dimensional $A_{2u}$, three-dimensional $T_{2u}$, and three-dimensional $T_{1u}$ for the orbital part of the wavefunction, and two-dimensional $\Gamma_6$ for the spin part. The 14-fold degenerate 4f orbital would be split into 2-fold $A_{2u}\otimes\Gamma_6$, 6-fold $T_{2u}\otimes\Gamma_6$, and 6-fold $T_{1u}\otimes\Gamma_6$ orbitals. Finite SOC splits further these orbitals and results in mixing of wavefunctions derived from different orbitals. The symmetry of the split states is given by the decomposition of the direct products into direct sums of IRs of O$_h$ double group~\cite{Koster1963}: $A_{2u}\otimes\Gamma_6\,=\,\Gamma_7$, $T_{2u}\otimes\Gamma_6\,=\,\Gamma_8\oplus\Gamma_7$, and $T_{1u}\otimes\Gamma_6\,=\,\Gamma_6\oplus\Gamma_8$.

On the other hand, if cubic CF were absent, the 4f orbital would be 
split into 8-fold $^2F_{7/2}$ ($J\,=\,L+S$) and 6-fold $^2F_{5/2}$ 
($J\,=\,L-S$) multiplets. Finite CF potential splits the two 
multiplets and induces mixing of wavefunctions derived from different 
multiplets~\cite{SM}.      
The symmetry of the split states is given by the compatibility table showing the mapping of IRs of the full rotational group into IRs of O$_h$ double group~\cite{Koster1963}: $^2F_{7/2}\,=\,\Gamma_8\oplus\Gamma_7\oplus\Gamma_6$, and $^2F_{5/2}\,=\,\Gamma_8\oplus\Gamma_7$. With both SOC and CF present, the CF eigenfunctions should be classified by the IRs of the double group, namely two-dimensional $\Gamma_6$, two-dimensional $\Gamma_7$, and four-dimensional $\Gamma_8$.

We diagonalize the Hamiltonian (\ref{eq:H}) in the basis of $|L,m_l\rangle|S,m_s\rangle$, where $L, m_l, S, m_s$ are quantum numbers corresponding to $\hat{\mathbf{L}}, \hat{L}_z, \hat{\mathbf{S}}, \hat{S}_z$ operators, respectively. After diagonalization, the CF transition energies can be expressed in terms of $\xi$, $B_4$ and $B_6$. We obtain these three parameters by fitting the energies of three CF transitions to the data at 15\,K (the weakest $\Gamma_8\rightarrow\Gamma_6^*$ transition is not accounted in this procedure). The obtained set of parameters comprises $\xi$=610\,cm$^{-1}$, $B_4$=-0.758\,cm$^{-1}$ and $B_6$=-0.0165\,cm$^{-1}$. The same values automatically render the energy of weakest transition at 2070\,cm$^{-1}$, which is close to the observed value at 2060\,cm$^{-1}$. The value of $\xi$ (610\,cm$^{-1}$) is also consistent with the estimated value for the Ce$^{3+}$ ion embedded in LuPO$_4$ (614\,cm$^{-1}$)~\cite{Williams1989}. Such consistency demonstrates the reliability of the model (\ref{eq:H}).

We can further use this single-ion model to describe the temperature dependence of the CF excitation energy. Here we assume that $\xi$ is temperature-independent, and that the temperature dependence of the $B_4$ and $B_6$ coefficients comes from the temperature dependence of the lattice constant $a(T)$. We therefore rewrite $B_4$ and $B_6$ as $B_4(T)$=$C_4a(T)^{-5}$ and $B_6(T)$=$C_6a(T)^{-7}$, where $C_4$ and $C_6$ are temperature-independent factors. The temperature dependence of the lattice constant $a(T)$ is obtained from the Refs.~\cite{Tanaka2002,Zaharko2003}. Then, we determine the values of $\xi$, $C_4$ and $C_6$ by matching the calculated values with the measured data at 300\,K. Finally, we use the determined $\xi$, $C_4$ and $C_6$ to calculate CF excitation energies below 300\,K. The results are shown in Fig.~\ref{fig:CF2}~(a-c). The discrepancy between the measured data and the calculated values below 200\,K results from unaccounted terms in the model Hamiltonian [Eq.~(\ref{eq:H})]; for an example, interactions between localized $f$-electrons and the itinerant conduction electrons.

By virtue of the obtained eigenfunctions, the Raman intensity of the four CF transitions can be calculated. For non-resonant scattering, the Raman response $\chi^{\prime\prime}(\omega)$ has the following expression~\cite{Devereaux2007}:
\begin{equation}
\chi^{\prime\prime}(\omega)\sim\frac{1}{Z}\sum_{i,f}|\langle f|\hat{R}_{\mu\nu}|i\rangle|^2e^{-E_i/kT}\delta(E_f-E_i-\hbar\omega)~,
\label{eq:I}
\end{equation}
where $Z$ is the partition function, $|i\rangle$, $|f\rangle$ are the initial and final state with energy $E_i$ and $E_f$, $\omega$ is the Raman shift, and $\hat{R}_{\mu\nu}$ is the effective Raman operator. In our case, $|i\rangle$ is the CF ground state and $|f\rangle$ is one of the excited CF states. For nonresonant Raman scattering, $\hat{R}_{\mu\nu}$ is a quadrupolar operator depending on the crystallographic symmetry and scattering geometry $\mu\nu$~\cite{Axe1964,Kiel1969,Williams1989}. For XY scattering geometry in a cubic crystal, $\hat{R}_{XY}$ transforms in the same way as quadrupole $xy$ under the symmetry operations of O$_h$ point group:
\begin{equation}
\hat{R}_{XY}=\frac{1}{2}(\hat{L}_x\hat{L}_y+\hat{L}_y\hat{L}_x)=\frac{1}{4i}(\hat{L}_+^2-\hat{L}_-^2)~,
\label{eq:R}
\end{equation}
where $\hat{L}_+$ and $\hat{L}_-$ are the ladder operators of the orbital angular momentum. We note that because light only couples to the electron's orbital degree of freedom, the effective Raman operator should be written in terms of the orbital angular momentum operators, rather than the total angular momentum operators. Expression~(\ref{eq:R}) should accordingly be evaluated in the basis of $|L,m_l\rangle|S,m_s\rangle$.

\begin{figure}
\includegraphics[width=0.48\textwidth]{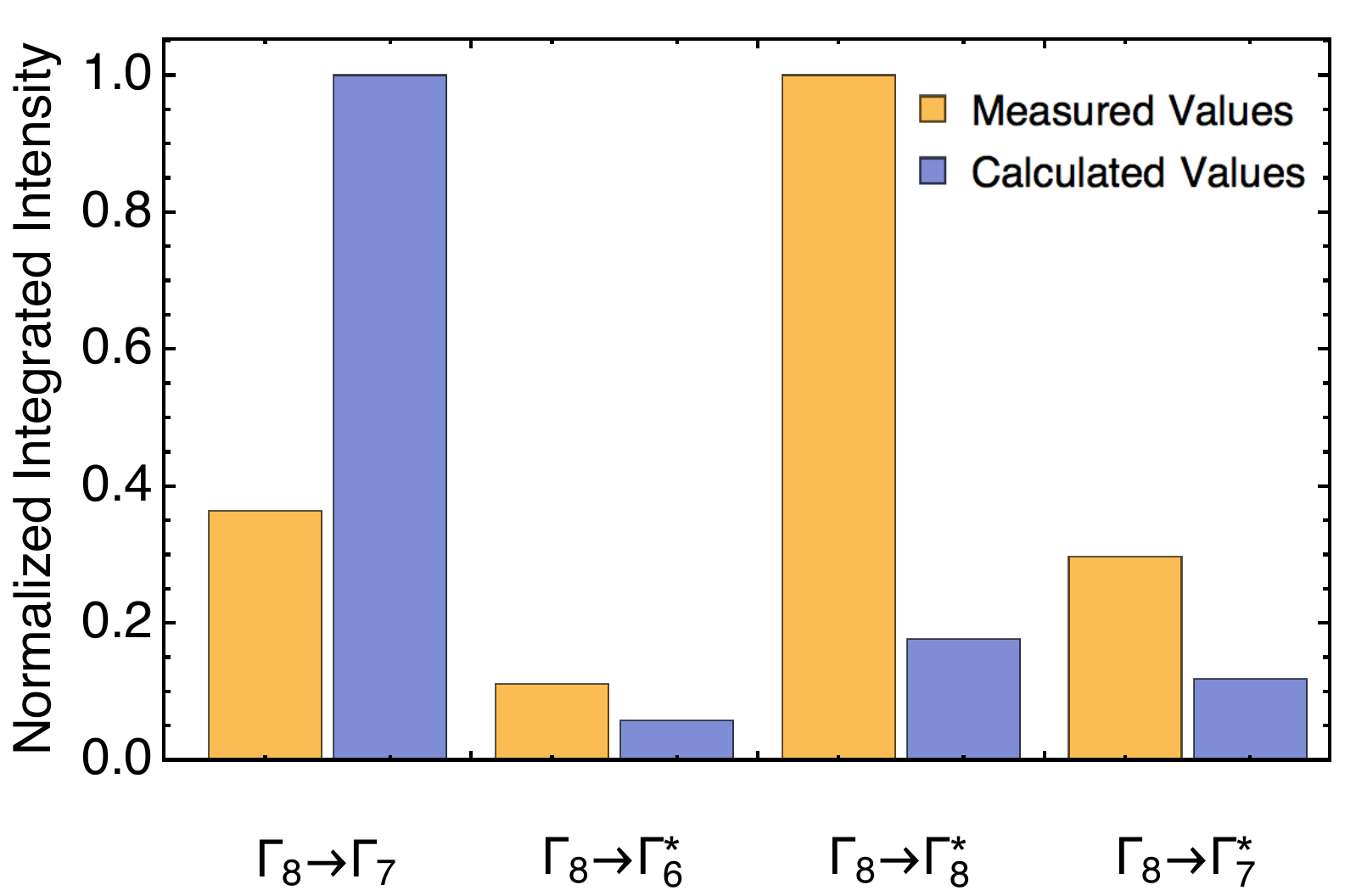}
\caption{\label{fig:CF5}Normalized intensity of the four CF transitions in XY scattering geometry at 15\,K, measured (in yellow) and calculated (in blue). The measured/calculated intensity of the four transitions is normalized by their respective largest value.}
\end{figure}

In Fig.~\ref{fig:CF5} we compare the calculated and measured CF transition intensity. Because the 476\,nm excitation is resonant with interband transitions (see SubSec.~\ref{subsec:PL}) but the expression~(\ref{eq:R}) is only valid for non-resonant scattering, we expect discrepancy between the calculated and measured results. Nevertheless, the relative intensity of the three inter-multiplet transitions is reproduced.

\subsection{Phononic Excitations\label{subsec:P}}

\begin{figure}
\includegraphics[width=0.48\textwidth]{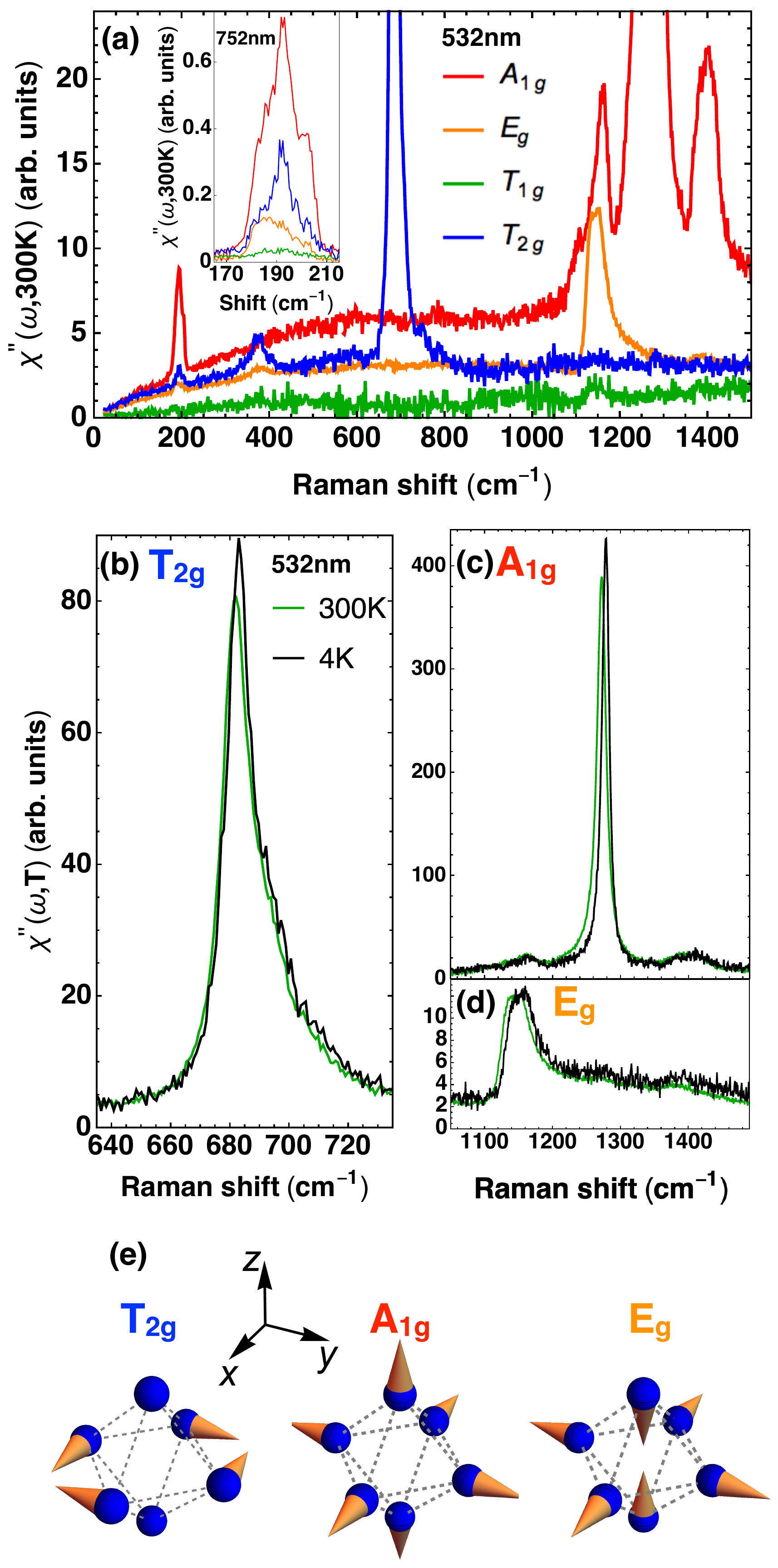}
\caption{\label{fig:P1}(a) Symmetry-decomposed Raman response $\chi^{\prime\prime}$($\omega$,300\,K), measured with 532\,nm excitation at 300\,K. Inset in (a): Symmetry-decomposed Raman spectrum of the second-order acoustic phonon scattering peak, measured with 752\,nm excitation at 300\,K. Thermal factor $[1+n(\omega/2,T)]^2$ is used to derive this particular inset; the other Raman spectra presented in this paper are obtained with the normal thermal factor $[1+n(\omega,T)]$. (b), (c) and (d): Raman spectra of the T$_{2g}$, A$_{1g}$ and E$_{g}$ optical phonon modes, measured with 532\,nm excitation at 300\,K and 4\,K. In (b), (c) and (d), the spectral resolution is 2.8\,cm$^{-1}$ for the high temperature data and 1.3\,cm$^{-1}$ for the low temperature data. (e) The schematic vibration patterns for the three optical phonon modes. Because the cerium ions are at the inversion centers, Raman-active phonon modes only involve vibrations of the boron octahedrons.}
\end{figure}

An overview of the phonon modes is presented in Fig.~\ref{fig:P1}(a). From group-theory analysis, CeB$_{6}$ has three Raman-active optical phonon modes: A$_{1g}$, E$_{g}$ and T$_{2g}$. Their respective energies are 1271, 1143 and 681.7\,cm$^{-1}$ at 300\,K, consistent with previous results~\cite{Zirngiebl1986,Ogita2003}. Their lineshapes at 300\,K and 4\,K are presented in Fig.~\ref{fig:P1}(b); no anomaly is observed on cooling. The E$_{g}$ and T$_{2g}$ optical phonon modes exhibits asymmetric lineshape. The underlying electronic continuum likely results from electronic interband transitions: according to the calculated and measured band structure~\cite{Kitamura1994,Suvasini1996,Neupane2015}, many direct interband transitions are allowed and in turn can contribute to the nearly flat continuum below 1500\,cm$^{-1}$ ($\sim$0.2\,eV).

The peak at 194\,cm$^{-1}$ is not fully polarized. It originates from second-order scattering of acoustic branches at the Brillouin-zone boundary~\cite{Ogita2003}, where the flat dispersion gives rise to a large density of states. From this peak, we infer that the maximum of the acoustic phonon frequency is around 100\,cm$^{-1}$, which is consistent with the INS data~\cite{Kunii1997}. Another feature at 373\,cm$^{-1}$ shows larger T$_{2g}$ contribution and smaller E$_{g}$ contribution. It is the $\Gamma_8\rightarrow\Gamma_7$ CF excitation discussed in SubSec.~\ref{subsec:CF}. The peak at 1400\,cm$^{-1}$ has strong A$_{1g}$ contribution and very weak E$_{g}$ contribution. It results from second-order scattering of the T$_{2g}$ phonon mode~\cite{Ogita2003}. The symmetry-decomposed spectra further reveal an A$_{1g}$ peak at 1158\,cm$^{-1}$, which was not reported previously. This peak might correspond to the summation mode of the 373\,cm$^{-1}$ CF excitation and the T$_{2g}$ phonon mode. Such coupling has been observed in another $f$-electron system UO$_2$~\cite{Livneh2008}.

\begin{figure}[b]
\includegraphics[width=0.48\textwidth]{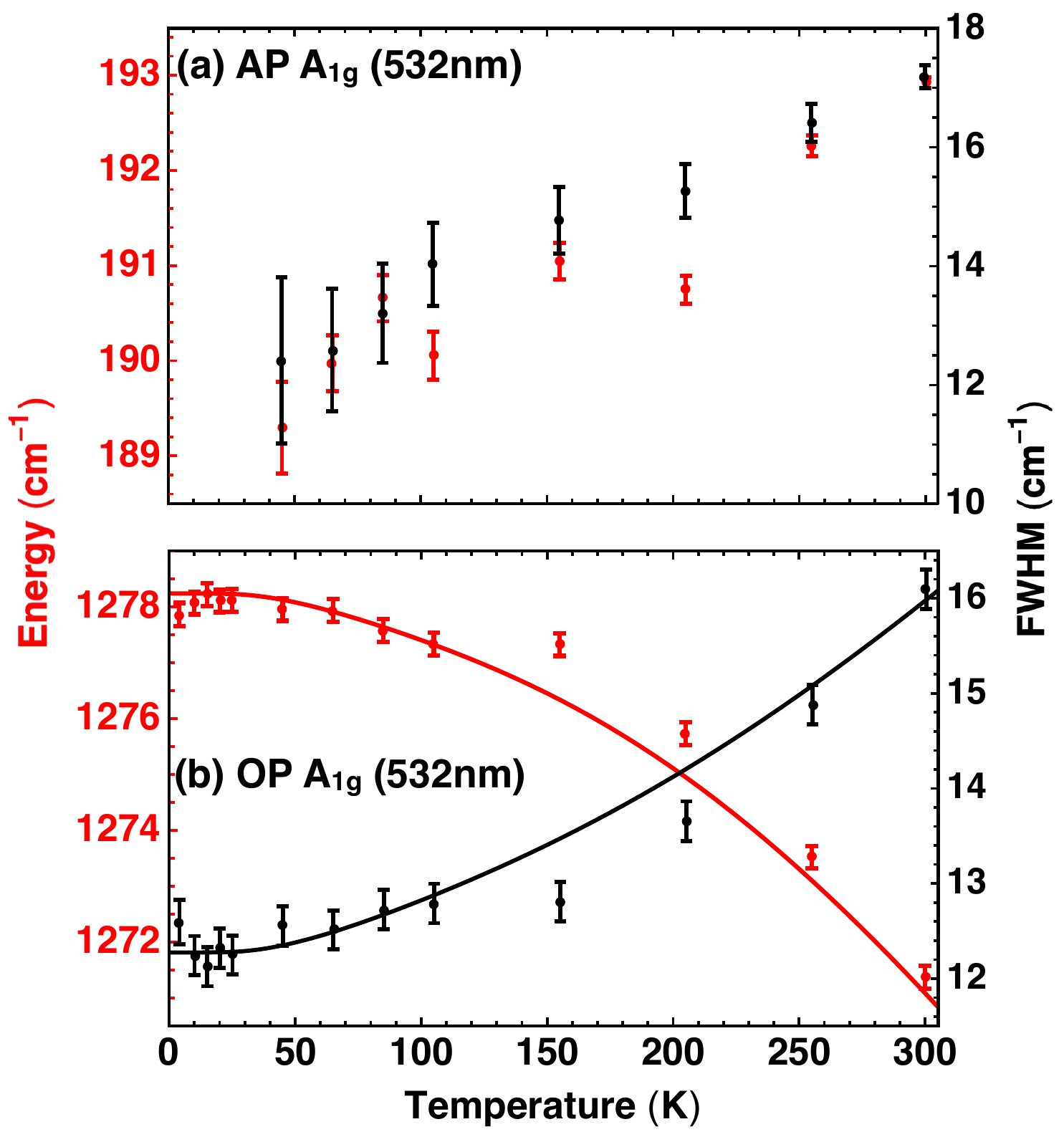}
\caption{\label{fig:P2} Temperature dependence of the energy (in red) and FWHM (in black) of (a) the A$_{1g}$ component of the second-order acoustic phonon scattering peak, and (b) the A$_{1g}$ optical phonon mode. The solid lines are fitting curves of an anharmonic decay model assuming decay into two optical modes, or an optical plus an acoustic modes~\cite{Wallis1966,Wallis1983}. The error bars represent one standard deviation of the Lorentzian fit.}
\end{figure}

In Fig.~\ref{fig:P2} we show the temperature dependence of the energy and FWHM of the A$_{1g}$ contribution of the second-order acoustic mode, and the A$_{1g}$ optical mode. The spectral parameters of the phonon modes were obtained by fitting the measured spectral peaks with a Lorentzian lineshape.

Temperature dependence of the phonon energy and FWHM is usually described by anharmonic effects. In most cases, the three-phonon processes renders the fastest relaxation, and higher-order processes can be neglected. Furthermore, the A$_{1g}$ optical mode at $\Gamma$ point has the highest frequency among all the phonon branches of CeB$_{6}$~\cite{Gurel2010}; hence we only need to consider processes in which one A$_{1g}$ optical mode at $\Gamma$ point decays into two phonon modes satisfying conservation of energy and momentum~\footnote{There are other constraints which the decay processes must satisfy. For example, the spontaneous decay of a phonon by anharmonic processes of any order into a set of phonons of higher phase velocity is impossible~\cite{Lax1981}.}. 
We use an generalized anharmonic decay model assuming multiple decay channels; for every channel, the decay products can be two acoustic modes, an optical plus an acoustic modes, or two acoustic modes~\cite{Wallis1966,Wallis1983}~\footnote{Because in CeB$_{6}$ the maximum acoustic phonon frequency is around 100\,cm$^{-1}$, for the high-frequency A$_{1g}$ mode decay into two acoustic modes is impossible.}:
\begin{equation}
\omega(T)=\omega_0-\sum_{i}\omega_{\delta (i)}[1+\frac{1}{e^{\hbar\omega_{1(i)}/k_BT}-1}+\frac{1}{e^{\hbar\omega_{2(i)}/k_BT}-1}]~,
\label{energyTwoDiff}
\end{equation}
\begin{equation}
\Gamma(T)=\Gamma_0+\sum_{i}\Gamma_{\delta (i)}[1+\frac{1}{e^{\hbar\omega_{1(i)}/k_BT}-1}+\frac{1}{e^{\hbar\omega_{2(i)}/k_BT}-1}]~,
\label{gammaTwoDiff}
\end{equation}
where the subscript $(i)$ indicates the decay channel. $\omega_{\delta (i)}$ and $\Gamma_{\delta (i)}$ are factors reflecting the relative importance of the various decay channels. $\hbar\omega_{1(i)}$ and $\hbar\omega_{2(i)}$ are the energy of the decay products in the decay channel labelled by $(i)$. $\hbar(\omega_0-\sum_{i}\omega_{\delta (i)})$ and $\Gamma_0+\sum_{i}\Gamma_{\delta (i)}$ 
correspond to the zero-temperature phonon energy and the FWHM, 
respectively. $\Gamma_0$ accounts for the temperature-independent part of the FWHM originating not from anharmonic decay processes, but from, for example, imperfection of the sample. 

Both $\omega_{\delta (i)}$ and $\Gamma_{\delta (i)}$ are proportional to
\begin{equation}
\sum_{\mathbf{k}_{1(i)},\mathbf{k}_{2(i)}}|\alpha(\mathbf{k}_{1(i)},\mathbf{k}_{2(i)})|^2\,\delta[\omega_{A_{1g}}-\omega_{1(i)}(\mathbf{k}_{1(i)})-\omega_{2(i)}(\mathbf{k}_{2(i)})]~,
\label{decayC}
\end{equation}
where $\alpha$ is the anharmonic coefficient; $\mathbf{k}_{1(i)}$ and $\mathbf{k}_{2(i)}$ are the wavevector of the decay products in the decay channel labelled by $(i)$; $\delta$ represents the Dirac $\delta$-function.

Referring to the calculated phonon dispersion~\cite{Gurel2010}, we expect two decay channels for the 1278\,cm$^{-1}$ A$_{1g}$ phonon: (1) decay into one 684\,cm$^{-1}$ optical phonon and one 594\,cm$^{-1}$ optical phonon with opposite momenta; (2) decay into one 1178\,cm$^{-1}$ optical phonon near +R point and one 100\,cm$^{-1}$ acoustic phonon near -R point.

The two phonon branches involved in the decay channel (1) is essentially flat over the whole Brillouin zone; hence a large number of states are available for the decay to happen. On the contrary, for the two phonon branches of the decay channel (2), only states near R point simultaneous satisfy the requirements of energy and momentum conservation. Therefore, the decay channel (1) would dominate if the anharmonic coefficient is not significantly different for the two channels.

\begin{table}[b]
\caption{\label{tab:P1}The fitting parameters for the energy and FWHM of the A$_{1g}$ optical phonon mode. Units are cm$^{-1}$.}
\begin{ruledtabular}
\begin{tabular}{ccc}
$\omega_0$&$\omega_{\delta (1)}$&$\omega_{\delta (2)}$\\
1309.0$\pm$0.1&28.12$\pm$0.05&2.664$\pm$0.003\\
\hline
$\Gamma_0$&$\Gamma_{\delta (1)}$&$\Gamma_{\delta (2)}$\\
1.07$\pm$0.08&8.4$\pm$0.2&1.76$\pm$0.01\\
\end{tabular}
\end{ruledtabular}
\end{table}

The fitting results of the anharmonic decay model are summarized in Table~\ref{tab:P1}. Indeed, $\omega_{\delta (1)}$\,\textgreater\,$\omega_{\delta (2)}$ and $\Gamma_{\delta (1)}$\,\textgreater\,$\Gamma_{\delta (2)}$. The temperature-independent $\Gamma_0$ is much smaller than $\Gamma_{\delta (1)}+\Gamma_{\delta (2)}$, indicating not only that the lineshape broadening mainly results from the anharmonic decay, but also that the sample is of excellent quality. In contrast to the behavior of the A$_{1g}$ optical mode, the second-order scattering of acoustic modes in the A$_{1g}$ channel shows decreasing energy on cooling [Fig.~\ref{fig:P2}(a)]. The 2\% softening might be a prelude to the AFQ ordering.

We attribute the apparent asymmetric lineshape of the T$_{2g}$ and 
E$_{g}$ optical phonon modes to the coupling between these phonons 
and the low-frequency fluctuations [SubSection.~\ref{subsec:QE}].   
The observed spectral lineshapes are resulted from convolution of the 
phononic Lorentzian and Drude-like function describing the low-lying 
fluctuations. 
We use the following expression to fit modes' lineshape at 4\,K:          
\begin{multline}
\chi^{\prime\prime}(\omega,4\,K)=
\sum_i\{\frac{A^2_{(i)}\gamma_{L(i)}}{(\omega-\omega_{L(i)})^2+\gamma_{L(i)}^2} +\\
\frac{A^2_{(i)}v_{(i)} \theta(\omega-\omega_{L(i)}) 
(\omega-\omega_{L(i)})[1+n(\omega-\omega_{L(i)},4\,K)]}{(\omega-\omega_{L(i)})^2+(\gamma_{L(i)}+\gamma_{D(i)})^2} +\\ 
\frac{A^2_{(i)}v_{(i)} \theta(\omega_{L(i)}-\omega) 
(\omega_{L(i)}-\omega) n(\omega_{L(i)}-\omega,4\,K)}{(\omega_{L(i)}-\omega)^2+(\gamma_{L(i)}+\gamma_{D(i)})^2}\}~.
\label{eq:Asym}
\end{multline}
In Eq.~(\ref{eq:Asym}), the first term describes the bare phonon 
part, while the second and third terms correspond to the Stokes and 
anti-Stokes of the phonon assisted electronic scattering. 
The summation runs over all 
the $k$-points in the Brillouin zone. 
Referring to the calculated phonon dispersion~\cite{Gurel2010}, the 
T$_{2g}$ mode belongs to a  
flat branch over the Brillouin zone, while the E$_{g}$ mode belongs a 
dispersive branch which has high DOS at $\Gamma$ and $R$ 
points~\cite{Gurel2010}. 
Therefore, for the latter case we only consider coupling at $\Gamma$ 
and $R$ points. 
In this equation, $A_{(i)}$ is the phonon light-scattering vertex; 
$\omega_{L(i)}$ is the phonon frequency; 
$2\gamma_{L(i)}$ is the FWHM of the bare phonon Lorentzian function; 
$\gamma_{Di}$ measures the relaxation rate of the Drude function; 
$v_{(i)}$ represents the electron-phonon coupling strength; 
and $\theta(\omega)$ is the Heaviside step function.                

For the T$_{2g}$ mode, we choose $\gamma_{D(\Gamma)}$ to be 
3.0\,cm$^{-1}$, which is consistent with the measured value of the 
T$_{1g}$ quasi-elastic fluctuations at 16\,K. 
For the E$_{g}$ mode, 
we choose both $\gamma_{D(\Gamma)}$ and $\gamma_{D(R)}$ to be 
11\,cm$^{-1}$, which is consistent with the measured value of the A$_{1g}$ 
quasi-elastic fluctuations at 16\,K. 
We further require that 
$v_{(\Gamma)}$ and $v_{(R)}$ are the same.       

\begin{figure}
\includegraphics[width=0.48\textwidth]{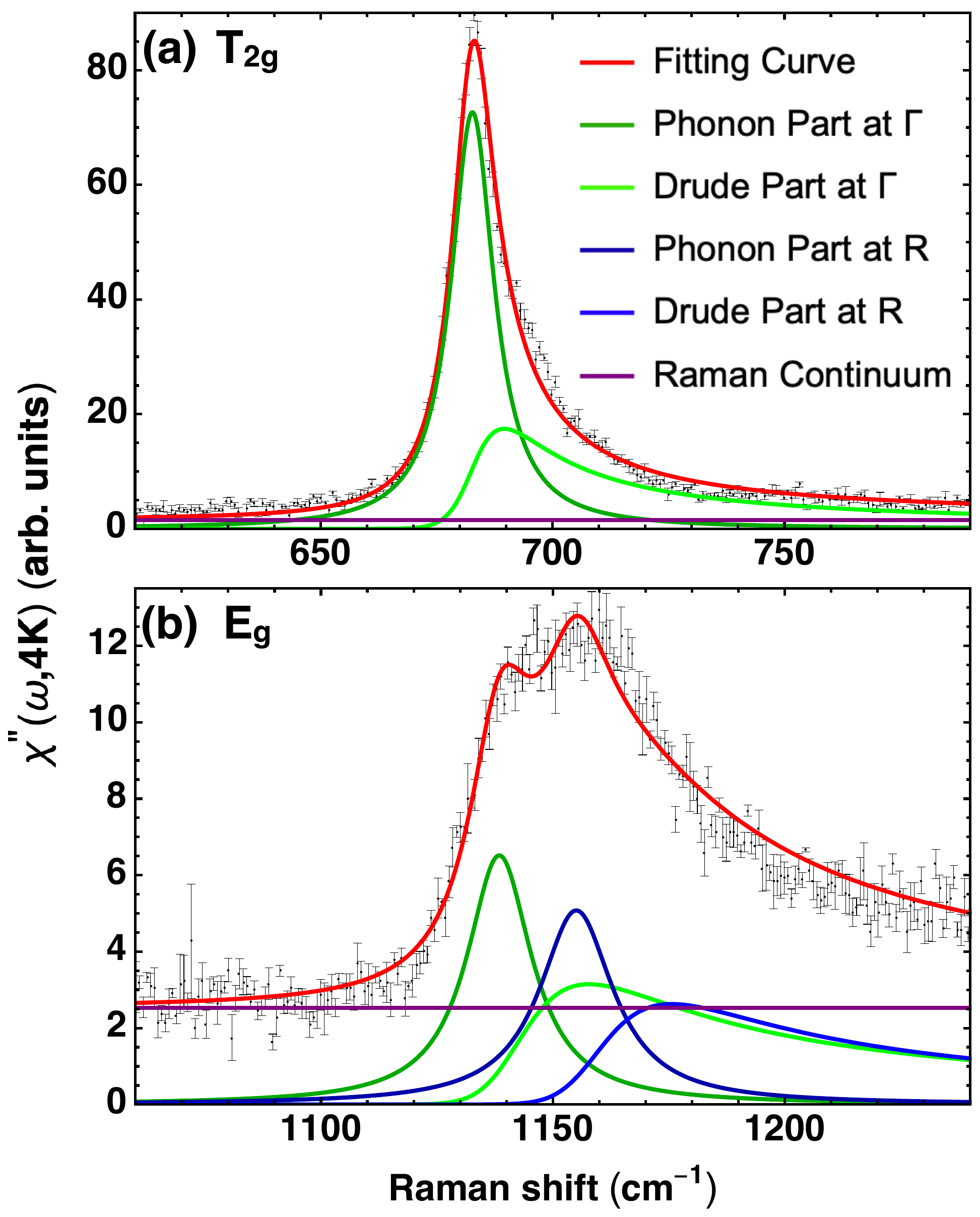}
\caption{\label{fig:Asym}
The measured with 532\,nm excitation at 4\,K Raman response function 
(black points with one standard deviation error bars) fitted with the model of Eq.~(\ref{eq:Asym}) for 
(a) the T$_{2g}$ and (b) the E$_{g}$ optical phonons coupled to 
low-frequency electronic excitations.}
\end{figure}

The fitting results of the T$_{2g}$ and E$_{g}$ composite modes are 
shown in Fig.~\ref{fig:Asym} and summarized in Table~\ref{tab:P2}. 
The dip of the fitting curve in Fig.~\ref{fig:Asym}(b) results from 
the negligence of the contributions at $k$-points between $\Gamma$ 
and $R$ points.
The FWHM of the bare T$_{2g}$ phonon mode 
($\sim$11\,cm$^{-1}$) is similar to that of the A$_{1g}$ phonon mode 
($\sim$12\,cm$^{-1}$), while the FWHM of the bare E$_{g}$ phonon mode 
($\sim$17\,cm$^{-1}$) is larger. 
This large E$_{g}$ FWHM, 
again, is an artifact caused by negligence of the contributions 
from remaining $k$-points. 
The energy difference between the E$_{g}$ mode 
at $\Gamma$ and $R$ points is $\sim$17\,cm$^{-1}$, which is 
comparable to the calculated difference of 
$\sim$30\,cm$^{-1}$~\cite{Gurel2010}.             

\begin{table}[b]
\caption{\label{tab:P2}The fitting parameters for the T$_{2g}$ and 
E$_{g}$ composite modes by Eq.~(\ref{eq:Asym}). Units are given 
in the brackets.}    
\begin{ruledtabular}
\begin{tabular}{lcc}
Parameter (Units)&T$_{2g}$ Mode&E$_{g}$ Mode\\
\hline
$A_{(\Gamma)}$ (a.u.)&20.29$\pm$0.03&7.4$\pm$0.4\\
$\gamma_{L(\Gamma)}$ (cm$^{-1}$)&5.67$\pm$0.02&8.5$\pm$0.5\\
$\omega_{L(\Gamma)}$ (cm$^{-1}$)&682.73$\pm$0.02&1138.4$\pm$0.3\\
$A_{(R)}$ (a.u.)&&7.0$\pm$0.8\\
$\gamma_{L(R)}$ (cm$^{-1}$)&&10$\pm$1\\
$\omega_{L(R)}$ (cm$^{-1}$)&&1155.0$\pm$0.5\\
$v$ (cm$^{-1}$)&0.691$\pm$0.005&2.2$\pm$0.2\\
\end{tabular}
\end{ruledtabular}
\end{table}

\subsection{Quasi-Elastic Excitations\label{subsec:QE}}

\begin{figure}[t]
\includegraphics[width=0.48\textwidth]{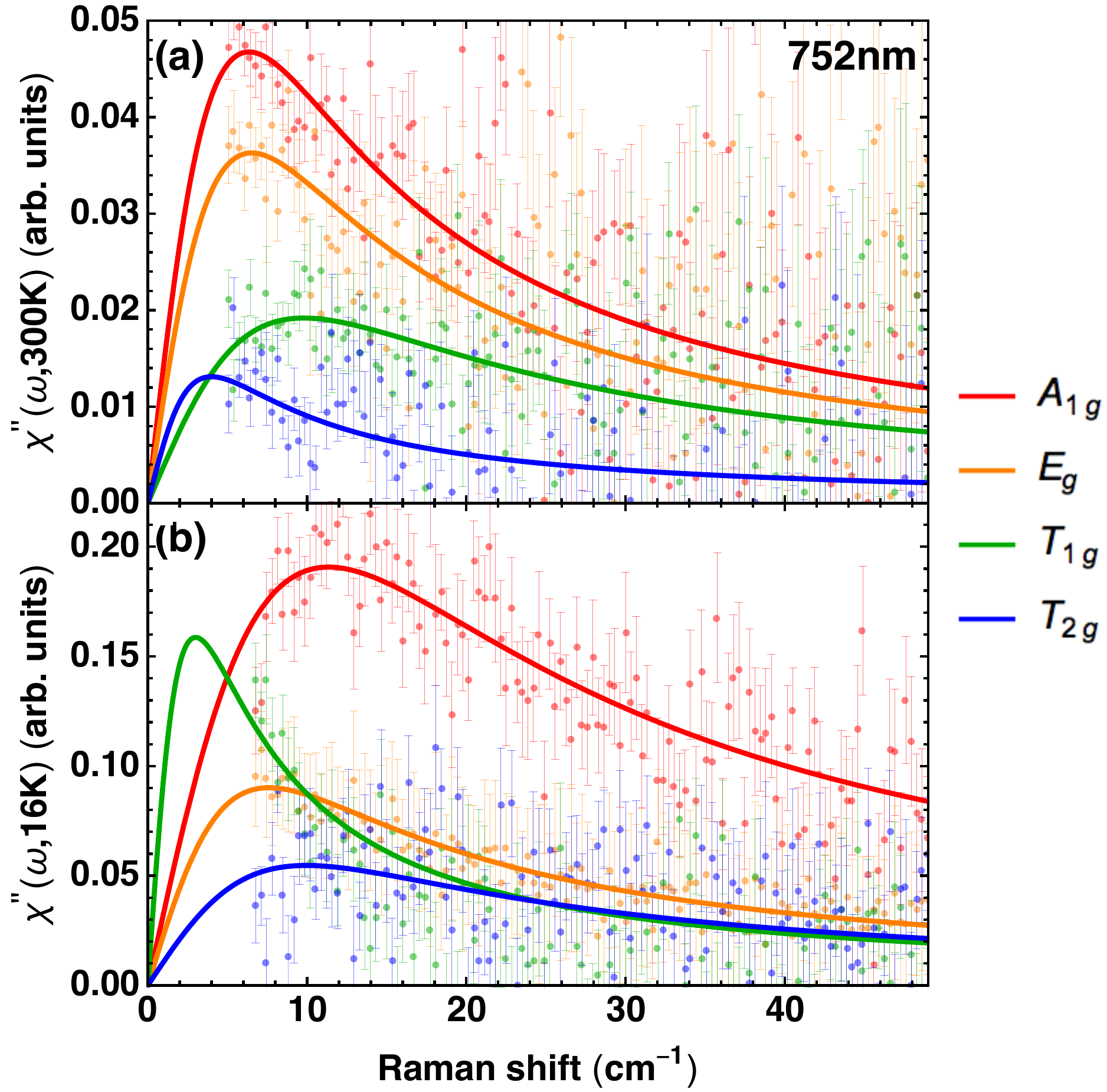}
\caption{\label{fig:QE1}Raman response $\chi^{\prime\prime}(\omega,T)$ in the four Raman-active symmetry channels measured with 752\,nm excitation at (a) 300\,K and (b) 16\,K. The solid lines are Drude fits [Eq.~(\ref{eq:Drude})]. The error bars represent one standard deviation.}
\end{figure}

In Fig.~\ref{fig:QE1} we show the symmetry-decomposed Raman response measured with 752\,nm excitation at 300\,K and 16\,K. The low-energy Raman response shows quasi-elastic features which can be described by a Drude lineshape:
\begin{equation}
\chi^{\prime\prime}(\omega,T)\propto\frac{\alpha^2\omega}{\omega^2+\gamma^2}~,
\label{eq:Drude}
\end{equation}
where $\alpha$ is the light-scattering vertex and $\gamma$ measures the fluctuation rate.

The Raman response gets enhanced in all the channels on cooling. Especially, the T$_{1g}$ Raman response changes qualitatively and develops into a strong quasi-elastic feature at low temperature. The basis functions of the T$_{1g}$ representation in O$_{h}$ group transform as the three components of angular momentum, which behave as a pseudovector~\cite{Koster1963}. This transformation property indicates that the observed quasi-elastic peak in T$_{1g}$ channel may have a magnetic origin.

\begin{figure}
\includegraphics[width=0.48\textwidth]{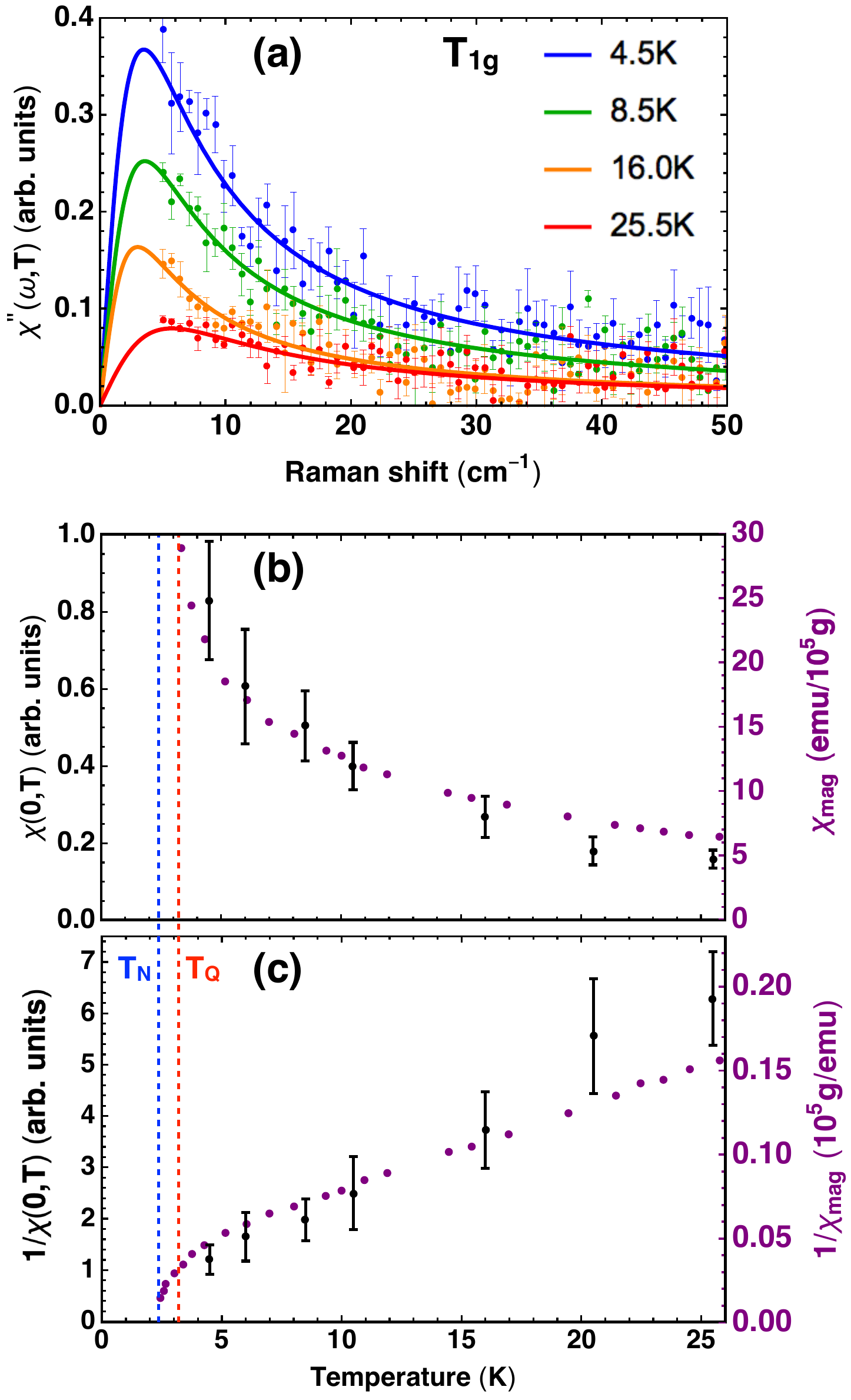}
\caption{\label{fig:QE2}(a) Temperature dependence of the Raman response $\chi^{\prime\prime}(\omega,T)$ in the T$_{1g}$ symmetry channel measured with 752\,nm excitation. The solid lines are Drude fits [Eq.~(\ref{eq:Drude})]. (b) Comparison between the temperature dependence of the static Raman susceptibility $\chi(0,T)$ (black) and that of the magnetic susceptibility $\chi_{mag}$ (purple)~\cite{Kawakami1980}. (c) Comparison between the temperature dependence of the inverse static Raman susceptibility (black) and that of the inverse magnetic susceptibility (purple)~\cite{Kawakami1980}. The blue arrow indicates the magnetic ordering temperature while the red one indicates the orbital ordering temperature. The error bars represent one standard deviation.}
\end{figure}

We measured the temperature dependence of Raman response in the XY 
scattering geometry, in which T$_{1g}$+T$_{2g}$ symmetry components 
are probed. Since T$_{2g}$ signal at low-temperature is nearly 
constant [Fig.~\ref{fig:QE1}(b)], we fit the Raman response with the 
sum of Drude and constant terms, and then remove the constant part to 
obtain the desired T$_{1g}$ component~\cite{SM}.
% ~\footnote{See Supplemental 
% Material for the justification of this procedure.}.          
The T$_{1g}$ Raman response obtained this way is shown in Fig.~\ref{fig:QE2}(a). The quasi-elastic excitation in T$_{1g}$ symmetry channel becomes significant below 20\,K, and its intensity increases on further cooling. The static Raman susceptibility, $\chi(0,T)$, plotted in Fig.~\ref{fig:QE2}(b) is obtained from the Raman response by virtue of Kramers-Kronig relation: $\chi(0,T)=\frac{2}{\pi}\int_{0}^{50\,cm^{-1}} \frac{\chi^{\prime\prime}(\omega,T)}{\omega} d\omega$. Drude function in Eq.~(\ref{eq:Drude}) is used to extrapolate $\chi^{\prime\prime}(\omega,T)$ below 4\,cm$^{-1}$. In Fig.~\ref{fig:QE2}(b) and (c), the temperature dependence of the static Raman susceptibility is compared with that of the magnetic susceptibility~\cite{Kawakami1980}. The fact that the temperature dependence of both quantities follows the same trend further supports the magnetic origin of the quasi-elastic peak in T$_{1g}$ symmetry channel~\cite{Kung2015}.

In zero magnetic field, Raman scattering data cannot determine whether the observed T$_{1g}$ quasi-elastic response is of FM or AFM origin. Nevertheless, the Raman-measured T$_{1g}$ quasi-elastic response is consistent with the FM correlations studied by INS: without external magnetic field and above T$_{Q}$, the magnitude of the INS-measured zone-center quasi-elastic peak decreases on warming~\cite{Jang2014}. We note by passing that a first-principle calculation for CeB$_{6}$ indicates that the expected values of both $4f$-orbital occupancy and total angular momentum exhibit an obvious anomalies around 20\,K~\cite{Lu2017}. This is the same temperature around which the T$_{1g}$ quasi-elastic Raman response starts to develop.

The mechanism responsible for the FM correlations can be understood as follows~\cite{Schlottmann2012}. Consider the two electrons at neighboring Ce$^{3+}$ sites. In the staggered orbital-ordering phase, the orbital part of the total wavefunction of these two electrons is antisymmetric. Due to the resulting exchange interaction, the spins at neighboring Ce$^{3+}$ sites are FM correlated.

The $\Gamma_8$ CF ground state of O$_{h}$ group has zero quadrupole moment. If the site symmetry is reduced from O$_{h}$ group to D$_{4h}$ group, the $\Gamma_8$ state of O$_{h}$ group would be split into the $\Gamma_6$ and $\Gamma_7$ states of D$_{4h}$ group. The $\Gamma_6$ and $\Gamma_7$ states can only have quadrupole moments of $x^2-z^2$ or $y^2-z^2$ type, rather than the proposed $xy$, $yz$, and $zx$ type. Hence, only when the site symmetry is reduced to D$_{2h}$ group, and the $\Gamma_8$ state of O$_{h}$ group is split into two $\Gamma_5$ states of D$_{2h}$ group, can the CF ground state carries finite quadrupole moments of $xy$, $yz$, and $zx$ type. However, in a continuous second-order phase transition, the symmetry of the system cannot be directly reduced from cubic to orthorhombic, which violates Landau theory~\cite{Landau1980}. Theories which claim an AFQ phase with O$_{xy}$-type moments using a localized picture should address this difficulty. Inconsistency of the AFQ description has also been suggested based on magnetic-susceptibility anisotropy and magnetostriction measurements~\cite{Amara2012}.

\section{Conclusion\label{sec:Con}}

In summary, we have employed optical secondary-emission spectroscopy to study the spin-orbital coupling (SOC), electronic crystal-field (CF) excitations, electron-phonon interaction and long-wavelength magnetic fluctuations in the heavy-fermion metal CeB$_{6}$.     

Ce$^{3+}$ ions have a single electron in the $4f$-shell. The SOC splits the degenerate 4f levels into a lower-energy $^2F_{5/2}$ multiplet and a higher-energy $^2F_{7/2}$ multiplet, with a separation of around 2000\,cm$^{-1}$, from which we estimate the SOC strength $\xi$=610\,cm$^{-1}$.

The two multiplets are further split into five Kramers-degenerate CF states by the cubic CF potential. The $^2F_{5/2}$ multiplet is composed of one quartet $\Gamma_8$ ground state and one doublet $\Gamma_7$ excited state, and the $^2F_{7/2}$ multiplet consists of $\Gamma_6^*$ and $\Gamma_7^*$ doublets, and a $\Gamma_8^*$ quartet states. We resolve all four electronic CF transitions: 380\,cm$^{-1}$ for the intra-multiplet excitation, and 2060, 2200 and 2720\,cm$^{-1}$ for the three inter-multiplet transitions. 

On cooling, the FWHM for the $\Gamma_8\rightarrow\Gamma_7$ and $\Gamma_8\rightarrow\Gamma_7^*$ transitions first decreases from 300\,K to 80\,K, but then increases below 80\,K. We relate the decrease of the FWHM to lattice vibration driven fluctuations of the electrostatic potential at Ce sites, which diminish on cooling. The increase of the FWHM below 80\,K results from the Kondo effect, an electron-correlation effect which increases the self-energy of the excited CF states.        
We apply a single-ion Hamiltonian model to obtain the eigenvalues and eigenfunctions of the 4f-electron CF states. Using the Fermi Golden Rule, we also calculate the intensity of the four Raman active CF transitions and compare the calculation to the experimental data.    

We study the lattice dynamics of CeB$_{6}$ and analyze the temperature dependence of all Raman active phonon modes. In the phonon spectra, we interpret the asymmetric lineshape of E$_{g}$ and T$_{2g}$ optical phonons as manifestation of electron-phonon interaction. We also identify a composite CF plus phonon excitation at 1158\,cm$^{-1}$. 

We acquire temperature dependence of the low-energy Raman response for all Raman-allowed symmetry channels, and uncover the development of a quasi-elastic Raman response in the magnetic-dipolar T$_{1g}$ symmetry channel below 20\,K. The corresponding static Raman susceptibility shows similar temperature dependence as the magnetic susceptibility data, which supports the interpretation of its magnetic origin. By comparing the quasi-elastic Raman scattering data with electron spin resonance and inelastic neutron scattering results, we relate this T$_{1g}$ spectral feature to ferromagnetic correlations.

Additionally, we detect photo-luminescence emission centered at 1.95\,eV at room temperature. We relate this emission to recombination of the electron-hole excitations between the 5d- and 4f-bands.

The experimental methods, models, and analyses demonstrated in this study can be applied to a range of systems, especially for rare-earth materials containing localized f-electrons of Ce$^{3+}$ or Yb$^{3+}$ ions at high-symmetry crystallographic sites~\cite{Ye2019}. The approach could enable us to probe ferroquadrupolar (FQ) fluctuations in TmAg$_{2}$ (T$_{FQ}$\,=\,5.0\,K)~\cite{Tm1993} or TmAu$_{2}$ (T$_{FQ}$\,=\,7.0\,K)~\cite{Tm1998} systems, to name a few examples. Also, magnetic correlation induced by quadrupolar ordering could be probed in antiferroquadrupolar (AFQ) systems, for instance in UPd$_{3}$ (multiple AFQ phases, with the highest T$_{AFQ}$\,=\,7.6\,K)~\cite{U1999}, NpO$_{2}$ (T$_{AFQ}$\,=\,25.0\,K)~\cite{Np2002}, or DyB$_{2}$C$_{2}$ (T$_{AFQ}$\,=\,24.7\,K)~\cite{Dy2000}.

\begin{acknowledgments}
We are grateful to K. Haule and P. Coleman for discussions. We thank 
A. Lee for participating in the early data acquisition. 
The spectroscopic work at Rutgers (M.Y., H.-H.K, G.B.) was supported 
by NSF Grant No. DMR-1709161.  
Sample synthesis at Los Alamos was performed under the auspices of 
the U.S. Department of Energy, Office of Basic Energy Sciences, 
Division of Materials Science and Engineering. 
G.B. also acknowledges 
the QuantEmX grant from ICAM, the Gordon and Betty Moore Foundation 
through Grant No. GBMF5305 allowing G.B. to make a collaborative 
visit to Stanford.      
Work at NICBP was supported by IUT23-3 grant. 
\end{acknowledgments}

%\bibliography{Ce.bib}


\newpage

\clearpage
\onecolumngrid
\appendix
% \setstretch{1.5}
\renewcommand{\thefigure}{S\arabic{figure}}
\addtocounter{figure}{-13}
\addtocounter{table}{-4}
\renewcommand{\theequation}{S\arabic{equation}}
\renewcommand{\thetable}{S\arabic{table}}
\begin{center}
    \textbf{\Large
Supplemental Material for: \\ 
Raman spectroscopy of $f$-electron metals: An example of CeB$_{6}$}
\end{center}

\section{Analysis of Raman Spectra}

\subsection{Subtraction of Photo-Luminescence}

\begin{figure}[b]
\includegraphics[width=0.44\textwidth]{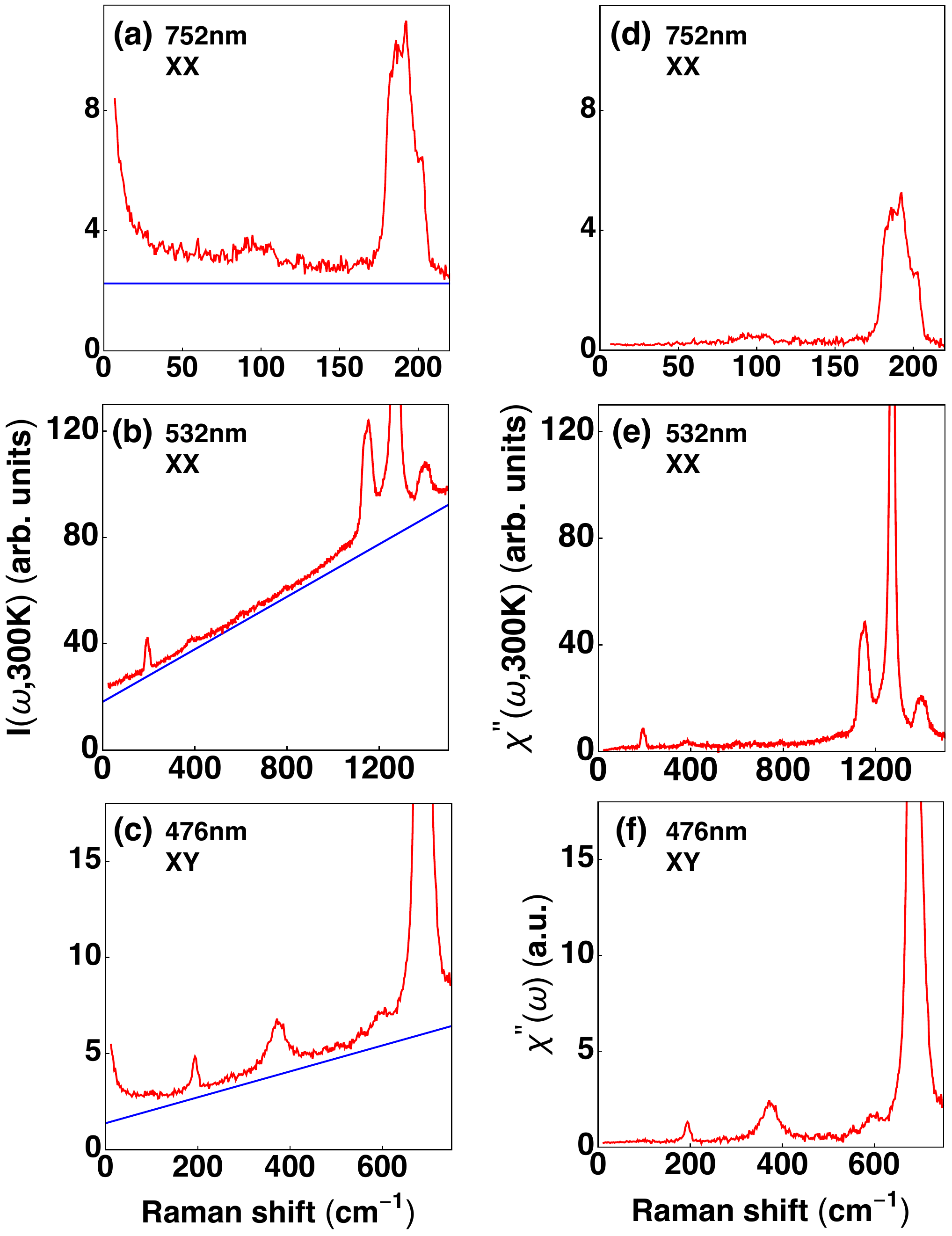}
\caption{\label{fig:SupPL}Illustration of subtracting photo-luminescence (PL) for spectra taken with three different excitation energy at 300\,K. Secondary-emission intensity $I(\omega)$ (red line) shown together with a constant PL (a) or a linear PL (b) and (c) (blue line) to be subtracted. The resulting Raman response $\chi^{\prime\prime}(\omega)$ is shown in (d), (e), and (f), respectively.}
\end{figure}

The measured secondary-emission intensity $I(\omega,T)$ is related to the Raman response $\chi''(\omega,T)$ by $I(\omega,T)=[1+n(\omega,T)]\chi''(\omega,T)+L(\omega,T)$, where n is the Bose factor, $\omega$ is energy, $T$ is temperature, and $L(\omega,T)$ represents photo-luminescence. Spectra taken with 752\,nm excitation are subtracted by a constant [Fig.~\ref{fig:SupPL} (a)], while those taken with 532 and 476\,nm excitations are subtracted by a linear function of frequency [Fig.~\ref{fig:SupPL} (b) and (c)].

\subsection{Thermal Factor for Second-Order Acoustic-Phonon Scattering}

For first-order scattering processes, the scattering intensity is given by the expression $[1+n(\omega,T)]\chi''(\omega,T)$~\cite{Hayes2004}, where $n$ is the Bose factor, $\chi''$ is the response function, $\omega$ is excitation energy and $T$ is temperature. However, for the second-order acoustic-phonon scattering process observed in our study, assuming the two constitute acoustic phonons have the same energy, the expression should be modified to $[1+n(\omega/2,T)]^2\chi''(\omega,T)$.

The reason for the modification is as follows. Second-order Raman scattering can result from either two successive first-order interactions, or one second-order interaction~\cite{Loudon1964}. In the first case, the thermal factor $[1+n(\omega/2,T)]$ should be used. However, it is essential that first-order scattering should be allowed for the two constituent excitations individually. Because the wavevector of visible light is much smaller than the Brillouin-zone size, first-order scattering of acoustic modes at the Brillouin-zone boundary is not allowed. The second-order Raman scattering of acoustic modes observed in our study, therefore, originates from the scattering process in which the light interacts with a pair of excitations in a single event. Wavevector conservation is effectively satisfied when the wavevectors of the constituent excitations are equal and opposite. In this case, assuming the two constituent excitations have the same energy, the thermal factor $[1+n(\omega/2,T)]^2$ should be used~\cite{Klein1981,Nyhus1997,Hayes2004}

\begin{figure}
\includegraphics[width=0.44\textwidth]{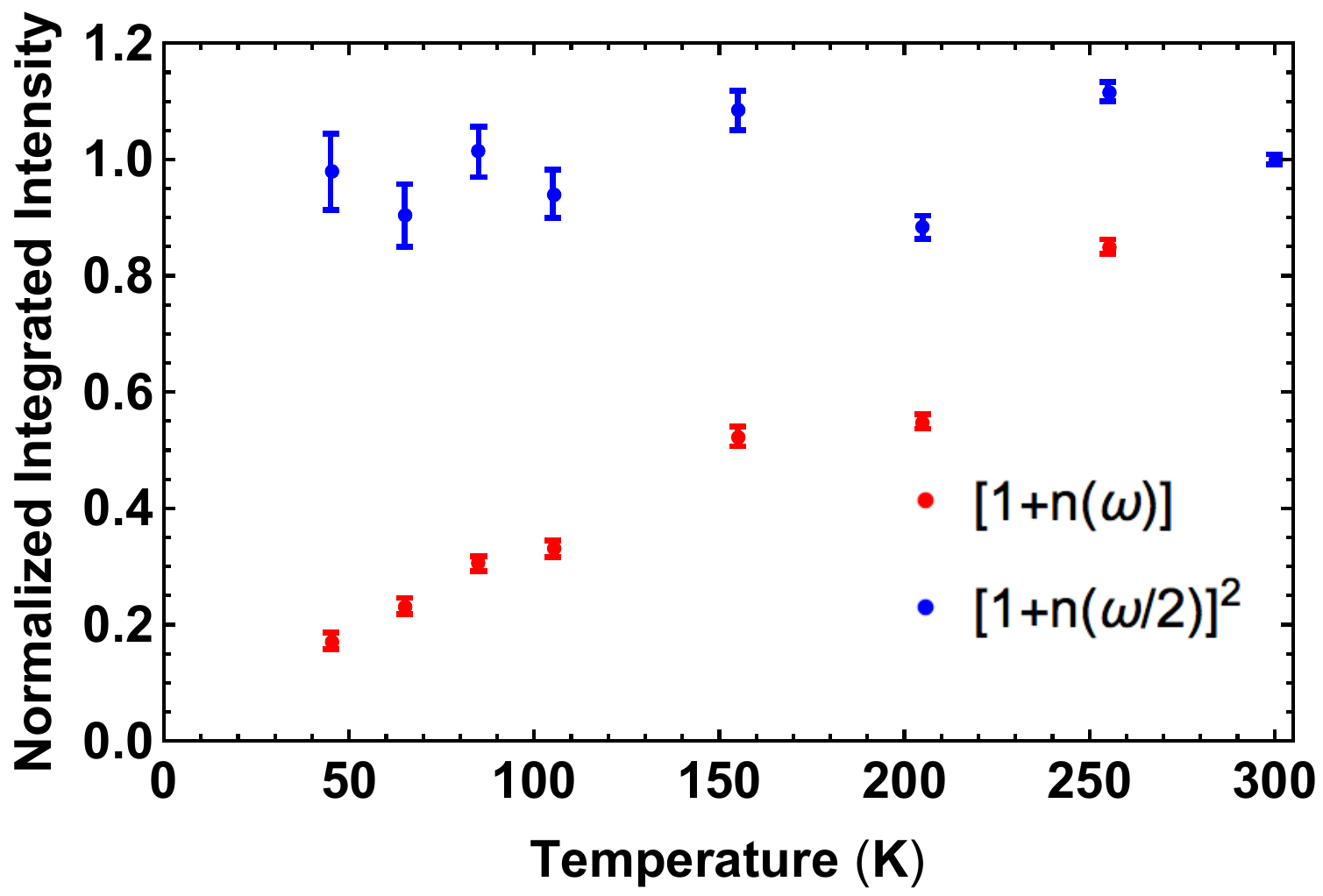}
\caption{\label{fig:SupAP}Temperature dependence of the integrated intensity of the A$_{1g}$ component of the second-order acoustic phonon scattering peak. The intensity obtained with two different thermal factors is normalized to their respective value at 300\,K. The error bars represent one standard deviation of the Lorentzian fit.}
\end{figure}

Using thermal factor $[1+n(\omega,T)]$ for the second-order acoustic-phonon scattering process in CeB$_{6}$ would lead to unreasonable decreasing intensity on cooling [Fig.~\ref{fig:SupAP}]. The energy and FWHM of the peak, on the contrary, is insensitive to which thermal factor is used.

\subsection{Measurement of T$_{1g}$ Quasi-Elastic Excitations}

\begin{figure}
\includegraphics[width=0.44\textwidth]{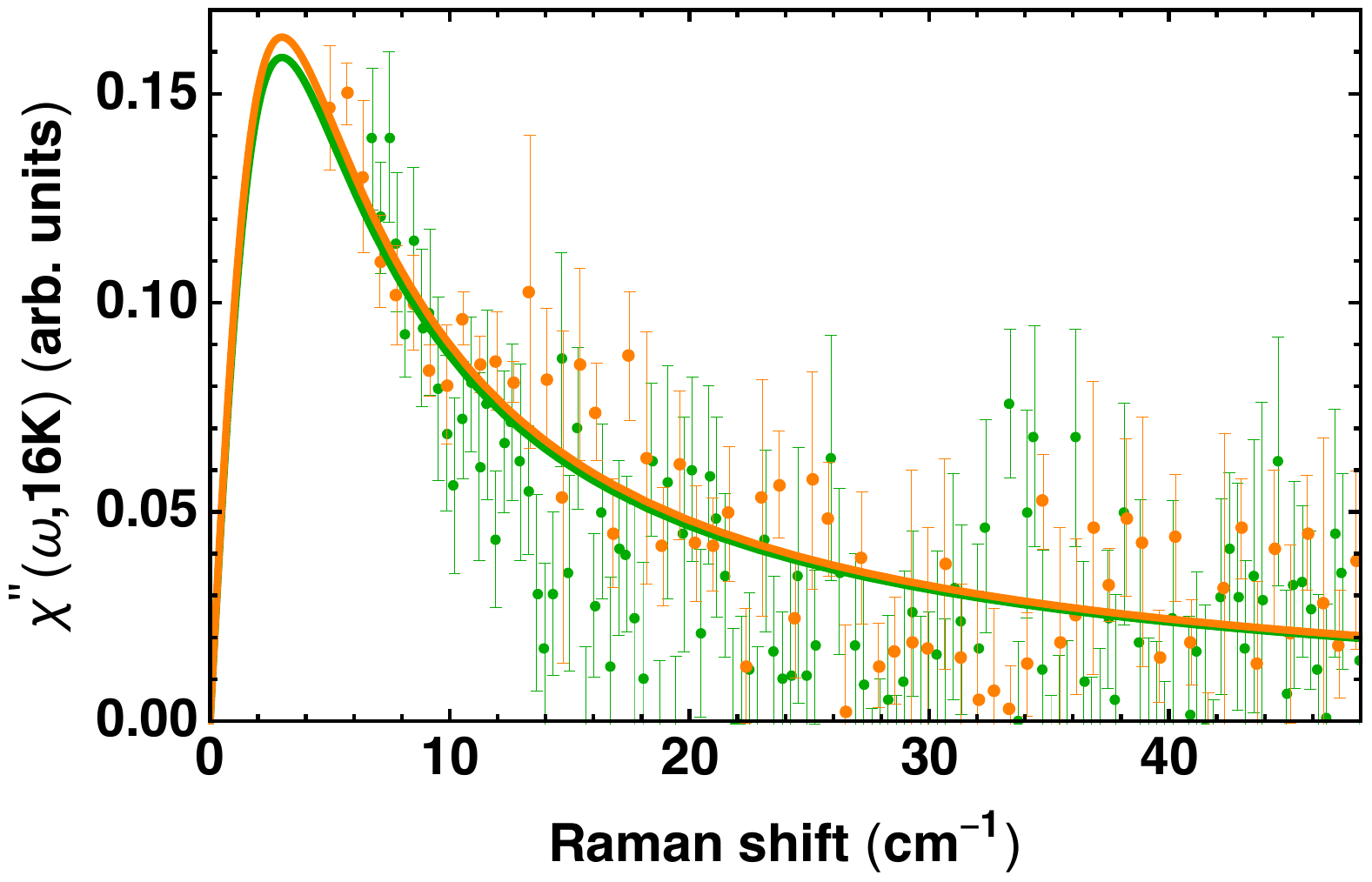}
\caption{\label{fig:SupQE}Comparison of the 16\,K T$_{1g}$ spectra in Fig.11(b) (green color) and Fig.12(a) (orange color). The solid lines are Drude fits. The error bars represent one standard deviation.}
\end{figure}

The T$_{1g}$ Raman response shown in Fig.12 of the Main Text is extracted from the data measured in the XY scattering geometry. We fit the Raman response with the sum of Drude and constant terms, and then remove the constant part to obtain the desired T$_{1g}$ component. To justify this procedure, Fig.~\ref{fig:SupQE} compares the 16\,K T$_{1g}$ spectrum shown in Fig.11(b) (decomposed from the spectra measured in the XY, X'Y' and RL scattering geometries) and in Fig.12(a) (obtained solely from the spectrum measured in the XY scattering geometry). They match well.

\section{Crystal-Field Eigenfunctions}

The calculated eigenfunctions are given in Table~\ref{table:F1}. Because of Kramers degeneracy, every eigenfunction $\sum_ic_i|L,m_l\rangle|S,m_s\rangle$ has a partner $\sum_ic_i|L,-m_l\rangle|S,-m_s\rangle$.

Usually, $H_{CF}$ is solved within a particular multiplet in the basis of $|J,m_j\rangle$~\cite{Fazekas1999}. This treatment assumes that SOC is much larger than CF potential, and inter-multiplet mixing can therefore be ignored. Without inter-multiplet mixing, the eigenfunctions are independent of the parameters $B_4$ and $B_6$. However, Ce has the smallest SOC among the 4f lanthanides. Thus comparing the exact results with the approximate results helps to illustrate the limitations of the approximate treatment. The eigenfunctions obtained by separately diagonalizing $H_{CF}$ in the $J=5/2$ and $J=7/2$ multiplets are presented in Table~\ref{table:F2}. For convenient comparison with Table~\ref{table:F1}, the results have been converted into $|L,m_l\rangle|S,m_s\rangle$ basis. Inter-multiplet mixing occurs between $\Gamma_{8(i)}$ and $\Gamma_{8(i)}^*$, and between $\Gamma_{7(j)}$ and $\Gamma_{7(j)}^*$, where $i=1,2,3,4$ and $j=1,2$. The absolute change of coefficients are larger for $\Gamma_{7}$ and $\Gamma_{7}^*$ states than for $\Gamma_{8}$ and $\Gamma_{8}^*$ states. $\Gamma_6^*$ state derived from the $J=7/2$ multiplet has no corresponding state in the $J=5/2$ multiplet; hence it has no inter-multiplet mixing, and the coefficients for $\Gamma_6^*$ state are the same in Table~\ref{table:F1} and~\ref{table:F2}. Notice that the approximate treatment not only changes the magnitude of various coefficients, but also makes some finite coefficients vanish.

\begingroup
\squeezetable
\begin{table*}
\caption{\label{table:F1}The eigenfunctions of the best-fit single-ion Hamiltonian for Ce$^{3+}$ in CeB$_{6}$. The coefficients of $|m_l,m_s\rangle$ are given; a blank entry means a zero coefficient.}
\begin{ruledtabular}
\begin{tabular}{lcccccccccccccc}
$m_s$&\multicolumn{7}{c|}{$-\frac{1}{2}$}&\multicolumn{7}{c}{$+\frac{1}{2}$}\\
$m_l$&-3&-2&-1&0&+1&+2&+3&-3&-2&-1&0&+1&+2&+3\\
\hline
$\Gamma_{8(1)}$  &-0.033 &       &       &       &-0.737 &       &       &       &       &       &+0.675 &       &       &       \\
$\Gamma_{8(2)}$  &       &       &       &+0.675 &       &       &       &       &       &-0.737 &       &       &       &-0.033 \\
$\Gamma_{8(3)}$  &       &-0.325 &       &       &       &-0.325 &       &+0.853 &       &       &       &+0.250 &       &       \\
$\Gamma_{8(4)}$  &       &       &+0.250 &       &       &       &+0.853 &       &-0.325 &       &       &       &-0.325 &       \\
$\Gamma_{7(1)}$  &       &-0.032 &       &       &       &+0.723 &       &+0.423 &       &       &       &-0.546 &       &       \\
$\Gamma_{7(2)}$  &       &       &-0.546 &       &       &       &+0.423 &       &+0.723 &       &       &       &-0.032 &       \\
$\Gamma_{8(1)}^*$&-0.763 &       &       &       &+0.455 &       &       &       &       &       &+0.459 &       &       &       \\
$\Gamma_{8(2)}^*$&       &       &       &+0.459 &       &       &       &       &       &+0.455 &       &       &       &-0.763 \\
$\Gamma_{8(3)}^*$&       &+0.477 &       &       &       &+0.477 &       &+0.152 &       &       &       &+0.722 &       &       \\
$\Gamma_{8(4)}^*$&       &       &+0.722 &       &       &       &+0.152 &       &+0.477 &       &       &       &+0.477 &       \\
$\Gamma_{7(1)}^*$&       &-0.816 &       &       &       &+0.380 &       &-0.267 &       &       &       &+0.345 &       &       \\
$\Gamma_{7(2)}^*$&       &       &+0.345 &       &       &       &-0.267 &       &+0.380 &       &       &       &-0.816 &       \\
$\Gamma_{6(1)}^*$&+0.645 &       &       &       &+0.500 &       &       &       &       &       &+0.577 &       &       &       \\
$\Gamma_{6(2)}^*$&       &       &       &+0.577 &       &       &       &       &       &+0.500 &       &       &       &+0.645 \\
\end{tabular}
\end{ruledtabular}
\end{table*}
\endgroup

\begingroup
\squeezetable
\begin{table*}
\caption{\label{table:F2}The eigenfunctions obtained by separately diagonalizing $H_{CF}$ in the $J=5/2$ and $J=7/2$ multiplets. The coefficients of $|m_l,m_s\rangle$ are given; a blank entry means a zero coefficient.}
\begin{ruledtabular}
\begin{tabular}{lcccccccccccccc}
$m_s$&\multicolumn{7}{c|}{$-\frac{1}{2}$}&\multicolumn{7}{c}{$+\frac{1}{2}$}\\
$m_l$&-3&-2&-1&0&+1&+2&+3&-3&-2&-1&0&+1&+2&+3\\
\hline
$\Gamma_8$  &       &       &       &       &-0.756 &       &       &       &       &       &+0.655 &       &       &       \\
$\Gamma_8$  &       &       &       &+0.655 &       &       &       &       &       &-0.756 &       &       &       &       \\
$\Gamma_8$  &       &-0.345 &       &       &       &-0.345 &       &+0.845 &       &       &       &+0.218 &       &       \\
$\Gamma_8$  &       &       &+0.218 &       &       &       &+0.845 &       &-0.345 &       &       &       &-0.345 &       \\
$\Gamma_7$  &       &-0.154 &       &       &       &+0.772 &       &+0.378 &       &       &       &-0.488 &       &       \\
$\Gamma_7$  &       &       &-0.488 &       &       &       &+0.378 &       &+0.772 &       &       &       &-0.154 &       \\
$\Gamma_8^*$&-0.764 &       &       &       &+0.423 &       &       &       &       &       &+0.488 &       &       &       \\
$\Gamma_8^*$&       &       &       &+0.488 &       &       &       &       &       &+0.423 &       &       &       &-0.764 \\
$\Gamma_8^*$&       &+0.463 &       &       &       &+0.463 &       &+0.189 &       &       &       &+0.732 &       &       \\
$\Gamma_8^*$&       &       &+0.732 &       &       &       &+0.189 &       &+0.463 &       &       &       &+0.463 &       \\
$\Gamma_7^*$&       &-0.802 &       &       &       &+0.267 &       &-0.327 &       &       &       &+0.423 &       &       \\
$\Gamma_7^*$&       &       &+0.423 &       &       &       &-0.327 &       &+0.267 &       &       &       &-0.802 &       \\
$\Gamma_6^*$&+0.645 &       &       &       &+0.500 &       &       &       &       &       &+0.577 &       &       &       \\
$\Gamma_6^*$&       &       &       &+0.577 &       &       &       &       &       &+0.500 &       &       &       &+0.645 \\
\end{tabular}
\end{ruledtabular}
\end{table*}
\endgroup

% \bibliography{Ce.bib}

\end{document}